\documentclass{article}

\usepackage{graphicx,amssymb}
\usepackage{graphics}
\usepackage{epsfig}
\usepackage{subfigure}
\usepackage{adjustbox}
\usepackage{amsmath}
\usepackage{setspace}

\usepackage{tikz}
\usepackage{multirow}
\usepackage{natbib,stfloats}
\usepackage{mathrsfs}

\begin{document}%

\setcounter{page}{1}

\title{Wakes in stratified atmospheric boundary layer flows: an LES investigation}

\author{John S. Haywood \& Adrian Sescu \vspace{3mm} \\
Department of Aerospace Engineering,\\ Mississippi State University,\\ Mississippi State, MS 39762, USA \\
{\it E-mail: jsh478@msstate.edu}}

\date{}

\maketitle

\begin{abstract}
Large eddy simulations and three-dimensional proper orthogonal decomposition were used to study the interaction between a large stationary and moving bluff body and a high Reynolds number stably-stratified turbulent boundary layer. An immersed boundary method is utilized to take into account the effect of the bluff body on the boundary layer flow and turbulent inflow conditions upstream of the bluff body are imposed by employing a concurrent precursor simulation method both because a pseudo-spectral method is utilized in the horizontal directions. The dominant POD mode type was observed to occur at higher energy levels in the more turbulent wakes. Varying the thermal stratification showed only a slight effect on the turbulent statistics, but a significant effect on the number of dominant POD modes at the highest energy levels.
\end{abstract}



%

\maketitle

\section{Introduction}

Dynamically significant flow structures that are developing around bluff bodies interacting with laminar or turbulent boundary layers can be captured by the use of large eddy simulation (LES), which offers a great potential over other lower order models, such as Reynolds Averaged Navier-Stokes Equations, and a reduced computational cost versus higher order models, such as direct numerical simulations. LES is based on the assumption that the large eddies in the flow are anisotropic and depend on the mean flow and the configuration geometry, while smaller eddies are isotropic and homogeneous and have a universal character. In LES, the instantaneous fluctuating flow field is split into resolved and subgrid scale (SGS) terms using a spatial filter of characteristic size which must be larger (or at least equal) to the numerical grid spacing. The equations are spatially filtered and flow structures that are larger than the filter size are explicitly predicted by the numerical solution to the Navier-Stokes equations, while the influence of smaller scales is accounted for via subgrid scale models.

Multiple investigations of different configurations of bluff bodies interacting with boundary layer flows have been performed over the years. In the early years, \cite{Murakami} demonstrated the potential of LES to predict a boundary layer flow around a wall-mounted cube, using a very coarse grid resolution given the computational resources of that time. Later on, \cite{Shah1,Shah2} performed new simulations of a boundary layer flow around bluff bodies and showed that LES is accurate enough in predicting various quantities of interest. In a review article, \cite{Rodi} compared LES and RANS predictions of channel flows around a cube and demonstrated the potential of LES to predict such complex flows accurately. 

In \cite{Nozawa}, an LES study was performed to quantify the mean and fluctuating surface pressures on a wall-mounted cube, with a precursor simulation used to generate appropriate boundary layer inlet conditions. They found good agreement with experimental data, which was measured in a boundary layer flow over a smooth surface. \cite{Ono} conducted an LES study of flows generated by a building that faced the incoming flow at 45 degrees. While, \cite{Zheng} used a combination of a RANS simulation with the Fluent inherent random flow generation to generate a turbulent boundary layer flow in order to compare the wind loads on a single tall building calculated from a LES study to wind tunnel experiments. \cite{Thomas} used LES of a surface-mounted cube in a turbulent boundary layer to study the effect of varying cube orientation on the surrounding flow. \cite{Sohankar} conducted an LES study of a square cylinder in uniform flow to compare different subgrid scale models.  In order to study the effects of varying anisotropy in turbulent flow on the mean flow around a square cylinder, \cite{Haque} compiled a library of grid-generated isotropic turbulence flow slices from a wind tunnel that were rescaled and used as inflow for the large eddy simulations. Focusing on the flow in the immediate vicinity of a bluff body, \cite{Yakhot} carried out a direct numerical simulation (DNS) of a surface-mounted cube in a fully-developed turbulent channel. With the goal of comparing to wind tunnel experiments, \cite{Lim} performed a wall-modeled LES study of a surface-mounted cube interacting with a turbulent boundary layer. \cite{DeStefano} used wall-resolved adaptive wavelet-collocation LES of a square cylinder in uniform flow to show that a variable threshold method is feasible for turbulent flows around bluff bodies. A grid resolution study was performed by \cite{Tseng} using a wall-modeled LES framework of both a square cylinder and a periodic matrix of cubes in uniform flow.  These results were then applied to particulate transport through a urban environment, which was modeled as a configuration of different sized boxes to resemble a cluster of buildings.  \cite{Bou-Zeid3} then conducted a building resolution study using wall-modeled LES to investigate the effect the level of building detail had on the surrounding neutral atmospheric boundary layer (ABL). 

An LES approach and the fully three-dimensional proper orthogonal decomposition (POD) technique are utilized to investigate how the dynamics of a stably-stratified turbulent boundary layer are affected by the presence of a large bluff body. The study is focused on the wakes that are the result of the interaction between the bluff body and the boundary layer flow. The numerical tool, consisting of a pseudo-spectral LES algorithm, solves the incompressible Navier-Stokes equations with the Boussinesq approximation in the momentum equations introduced to take into account the buoyancy effect. The effect of the body on the boundary layer flow is taken into account via an immersed boundary method. One of the main challenges of the study is the imposition of a realistic turbulent conditions in the upstream of the body, and this is achieved by utilizing a concurrent precursor simulation performed in a separate flow domain; a blending region is introduced at the end of the main simulation to transfer the data from the precursor simulation. This study is one of the first attempts to characterize the wake generated by the bluff body in high Reynolds number turbulent boundary layer by a fully three-dimensional POD technique. Whereas the body height Reynolds number is on the order of $10^4$ to $10^5$ in the above studies, this work considers a Reynolds number on the order of $10^7$. Results consisting of instantaneous velocity, potential temperature, and POD modes for streamwise and wall-normal velocity components contour plots  along with vertical profiles of vertical turbulent momentum and heat fluxes and turbulence intensity are reported and discussed.

\section{LES framework}

The governing equations employed are the filtered momentum, potential temperature transport, and continuity equations for incompressible flow. Included in the momentum equations is the Boussinesq approximation term.
\begin{eqnarray}\label{eq1}
\frac{\partial \tilde{u}_{i}}{\partial t}
+ \tilde{u}_{j}\frac{\partial \tilde{u}_{i}}{\partial x_{j}}
=
- \frac{\partial \tilde{p}^{*}}{\partial x_{i}}
- \frac{\partial \tilde{\tau}_{ij}}{\partial x_{j}}
+ \delta_{i3} g \frac{\tilde{\theta} - \langle \tilde{\theta} \rangle }{\theta_{0}}
+ F_{i}
\end{eqnarray}
\begin{eqnarray}
\frac{\partial \tilde{\theta}}{\partial t}
+ \tilde{u}_{j}\frac{\partial \tilde{\theta}}{\partial x_{j}}
=
- \frac{\partial \pi_{j}}{\partial x_{i}}, 
\hspace{12mm}
\frac{\partial \tilde{u}_{i}}{\partial x_{i}} = 0 
\end{eqnarray}
$\tilde{u}_{i}, i=1,2,3$ represents the velocity components along the axial, spanwise, and vertical ($x_1$-,$x_2$-,$x_3$-) directions, respectively. The potential temperature is $\tilde{\theta}$, while $\theta_{0}$ is a characteristic potential temperature. The bracketed angles represent a horizontal averaging, $g$ represents the gravitational acceleration, and the Kronecker delta is given by $\delta_{ij}$. The effective pressure, which is divided by density, is represented by $\tilde{p}^{*}$. $F_i$ is a forcing term. A Lagrangian scale-dependent model, developed by \cite{Bou-Zeid2} and later extended by \cite{Porte-Agel2} to scalar transport, is used to model the subgrid scale (SGS) stress, $\tau_{ij}$, and the SGS heat flux, $\pi_{j}$.

Because the Reynolds number associated with the ABL is very large, the flow in the proximity to the ground is modeled using Monin-Obukhov similarity theory (\cite{Monin}), and the molecular viscous diffusion term in the momentum equation is neglected because it is assumed to be very small. The presence of a bluff body in the boundary layer flow is modeled using the direct forcing immersed boundary method (IBM) first introduced by \cite{Mohd-Yusof}. In the direct forcing approach, the velocity boundary conditions are imposed by a force added to the discretized momentum equations.
\begin{equation}
\frac{u^{n+1}_i - u^n_i}{\Delta t} = RHS_i + f_i
\end{equation}
The imposed force can then be calculated by specifying the velocity at the boundary $(u_i)_{box}$.
\[
 f_i = 
 \begin{cases}
  \frac{(u_i)_{box} - u^n_i}{\Delta t} - RHS_i &  ;\ box\ points  \\
  0 & ;\ fluid\ points
 \end{cases}
\]
Through substituting the imposed force into the discretized momentum equations, it is shown that the boundary condition,$u^{n+1}_i = (u_i)_{box}$, is satisfied at each discrete point.

Under the assumption that the turbulence in a ABL is horizontally-homogeneous, periodic boundary conditions are used in the horizontal directions. However, here a concurrent precursor simulation provides inflow boundary conditions that are blended at the end of the domain, while keeping the periodicity condition in the streamwise direction. The vertical gradients of velocity and scalars and the vertical component of velocity are set to zero at the top boundary. The top of the domain is located well above the height of the boundary layer. The wall stress is related to the horizontal velocity components at the first grid point via the Monin-Obukhov similarity theory at the bottom boundary (\cite{Businger}). While the similarity theory was developed for averaged quantities, \cite{Moeng} imposed the wall stress in a strictly local sense. \cite{Bou-Zeid1,Bou-Zeid2} then showed that by using local filtered velocities, the local similarity formulation produces an average stress that is very close to the averaged similarity formulation.
\begin{eqnarray}
\tau_{i3}\vert_{z=0} = -u_{*}^2 \frac{\tilde{u_i}}{V_f} =
-\left(
\frac{\kappa V_f}
     {\ln \left( \frac{z}{z_{0}} \right) - \Psi_M}
\right)^2
 \frac{\tilde{u_i}}{V_f};\\ 
 i = 1,2 \nonumber
\end{eqnarray}
The wall stress components are $\tau_{13}\vert_{z=0}$ and $\tau_{23}\vert_{z=0}$ and the friction velocity is $u_{*}$. $\kappa$ is the von K\'{a}rm\'{a}n constant (taken to be $\kappa = 0.4$) and $z_{0}$ is the effective roughness length. $\Psi_M$ provides a stability correction  and $V_f = \left[ \tilde{u_1}(\Delta z/2)^2 + \tilde{u_2}(\Delta z/2)^2 \right]^{1/2}$ is the filtered horizontal velocity at the first grid point. Within the same theory, the heat flux at the surface is computed as
\begin{eqnarray}
\langle w' \theta' \rangle_{z=0} =
\frac{u_{*} \kappa \left( \theta_s - \tilde{\theta} \right)}
     {\ln \left( \frac{z}{z_{0s}} \right) - \Psi_H}
\end{eqnarray}
where $\theta_s$ is the imposed potential temperature at the surface, $\tilde{\theta}$ is the resolved potential temperature in the first grid point, $z_{0s}$ is the roughness length for scalars, $\Psi_H(\zeta)=\int_{0}^{\zeta}[1-\phi_H(\zeta')d\zeta'/\zeta']$ provides a  stability correction where $\zeta = z/L$ and $
\phi_H \left( \zeta \right) = 
Pr_t + \beta \zeta,
$
for stable conditions (here, the Prandtl number is $Pr_t=0.74$, and $\beta \approx 5$). An artificial heat source or sink, developed by Sescu and Meneveau \cite{Sescu}, is added above the top of the ABL within the precursor simulation to enable the desired atmospheric stability. 

The numerical method implemented in a pseudo-spectral LES code uses a pseudo-spectral horizontal discretization and a centered finite difference vertical discretization (\cite{Calaf1}, \cite{sescu2}). The aliasing errors introduced through the handling of nonlinear terms by the pseudo-spectral method are corrected according to the $3/2$ rule (\cite{Canuto}). The continuity equation is satisfied through the solution of the Poisson equation resulting from taking the divergence of the momentum equation. In the vertical direction, a staggered grid is used to avoid decoupling the velocity and pressure. This is implemented by storing the vertical velocity component at a location halfway between the location of the other dependent quantities. The solution is marched in time using a fully-explicit second-order Adams-Bashforth scheme (\cite{Butcher}).

\section{Concurrent precursor simulation}\label{}

Turbulent inflow conditions at the inflow boundary can be imposed using a precursor simulation, wherein a library of turbulent data is generated by a separate simulation performed prior to the running of the main simulation (\cite{Tabor}). The library of turbulent data is created in a periodic domain by saving a slice of the flow variables normal to the streamwise direction at every time step to disk. These slices are imposed as inflow boundary conditions for the main flow simulation. In this way the main simulation is provided with a realistic turbulent velocity field and scalars at the inlet (for example, \cite{Park}, \cite{Wu1,Wu2}). The memory that is necessary to store all of the turbulent data and the time used to read and impose the turbulent slices at the inlet of the main simulations increases as the number of grid points increases (\cite{Dhamankar}). 

Another approach is to run the precursor and main simulations simultaneously; this method was first introduced by \cite{Lund} in the context of turbulent flat plate boundary layers and then later improved upon by \cite{Ferrante}. In the method of Lund et al., a slice of flow data taken at a streamwise location of the precursor simulation is rescaled and imposed at the inlet of the main simulation. \cite{Mayor} used a recycling region upstream of the domain where perturbations at the end of the recycling region are added to a mean flow from a previous simulation. The outlet of the recycling region serves as the inflow conditions for the domain and the combination of the perturbations and the previous mean flow serve as the inflow for the recycling region.

\begin{figure}
 \begin{center}
        \mbox{
        \subfigure {\label{Pre-Main_a}\includegraphics[width=12cm]{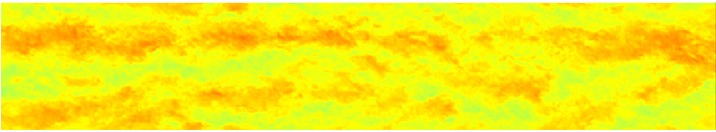}}    
         }
        \mbox{
        \subfigure {\label{Pre-Main_b}\includegraphics[width=12cm]{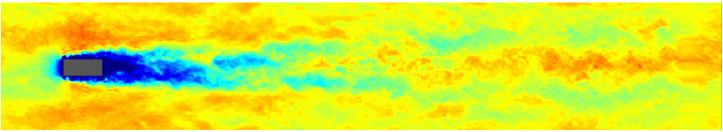}}
         }
 \end{center}
  \caption{Instantaneous streamwise velocity contour on a horizontal plane sectioning the box through the center: top) precursor domain; bottom) main domain.}
  \label{Pre-Main}
\end{figure}

Consisted with the approach proposed by \cite{Stevens}, the precursor and main flow domains considered here are identical, except the bluff body is added to the main domain. To avoid time interpolation the two simulations are synchronized in time. After each time step, a region of the flow data near the outlet of the precursor domain is blended on to the flow data in a region located in the proximity of the outflow boundary of the main domain. The blending region ensures that the flow data is smoothly transferred from the precursor simulation to the main simulation. Since the precursor domain and the main domain are identical, except for the presence of the body in the main domain, the Stevens method is capable of providing accurate turbulent inflow conditions. However, since the two domain have similar sizes, there is an increase in computational time and memory compared to other techniques, such as those based on recycling and rescaling. Assuming that the length of the blending region is $L_{blend}$ and ranges from $x = L_s$ to $x = L_x$, where $L_x$ is the length of entire domain, a generic flow variable (velocity or temperature) in the blending region can be calculated according to
\begin{eqnarray}
\phi^{blend}_i(x,y,z,t) =\ &w(x)&\phi^{pre}_i(x,y,z,t) \\ &+&\left[1-w(x)\right]\phi^{main}_i(L_s,y,z,t) \nonumber
\end{eqnarray}
where $\phi^{pre}_i$ and $\phi^{main}_i$ are flow variables from the precursor and main domains, respectively. The blending function used is
\[ 
w(x) = 
  \begin{cases}
    \frac{1}{2}\left[1-cos\left(\pi \frac{x-L_s}{L_{pl}-L_s}\right)\right] &;\ L_s \leq x \leq L_{pl} \\
    1  & ;\ L_{pl}<x \leq L_x
  \end{cases}
\]
where $L_{pl} = L_x - \frac{1}{4} L_{blend}$. Figure \ref{Pre-Main} shows an example involving an LES of a turbulent boundary layer with inflow conditions obtained from a concurrent precursor simulation; contours of the streamwise velocity are plotted for the precursor and main simulations. The transition inside the blending region can be seen at the end of the main simulation in figure \ref{Pre-Main_b}.


\section{Proper orthogonal decomposition}\label{}

Proper orthogonal decomposition (POD) is a modal decomposition technique used to determine the highest energy associated with coherent structures in a given flow.  In a typical POD analysis, the flow is decomposed into orthogonal modes, with the corresponding eigenvalues representing contributions of the modes to the flow turbulent kinetic energy. \cite{Lumley} was the first to apply POD analysis to turbulent flows. Later, the snapshot POD analysis was introduced by \cite{Sirovich} to deal with large flow data resulting from three-dimensional simulations. As examples of previous studies, \cite{Huang} used a two-dimensional POD analysis to study the coherent structures in ABL, and \cite{Liu} analyzed the vertical velocity profiles of the ABL using one-dimensional snapshot POD method to investigate mesoscale thermal structures. \cite{Shah} simulated the ABL over a range of stability conditions and used a two-dimensional snapshot POD method to look at the effect buoyancy had on the very large scale flow structures. \cite{Bastine} used the two-dimensional snapshot POD analysis of a single wind turbine wake to work towards a simplified dynamic wake model. Recently, \cite{VerHulst} employed a three-dimensional snapshot POD method to study the flow within large wind farms; this framework is utilized in the current work. For an in-depth overview of modern modal decomposition methods the reader is advised to read the recent review article by \cite{Taira} (where various tools are discussed and presented, and strengths and weaknesses of each modal decomposition method are discussed).

In the snapshot POD analysis, the decomposition of the velocity field is defined as 
\begin{equation}
u_i(x,y,z,t) = \overline{u}_i(x,y,z)+ \sum_{k=1}^{N} a^k(t) \psi^k_i(x,y,z),
\end{equation}
where $\overline{u}_i$ is the time averaged velocity, the number of snapshots is represented by $N$, the time-coefficients are $a^k$, and $\psi^k_i$ are the POD modes. The POD modes in the decomposition are orthonormal to each other, that is,
$
\langle \psi^k_i \psi^l_i \rangle_{xyz} = \delta_{kl}
$,
and the time-coefficients satisfy the relationship 
$
\overline{a^ka^l} = \lambda^k \delta_{kl}
$,
where $\lambda^k$ represents the strength of the POD mode $k$. The contribution of the POD mode $k$ to the time averaged turbulent kinetic energy is represented by $\lambda^k$.

Snapshots of the velocity field are saved every $\tau$ which is a time scale of the same order as the correlation time. The definition of the  $n^{th}$ snapshot is 
\begin{equation}
u_i(x,y,z,t_n) = u_i(x,y,z,n \tau)
\end{equation}
and it is assumed that none of the snapshots are correlated with each other. The fluctuating velocity field associated with each snapshot can be determined by subtracting out the mean velocity.
\begin{equation}
u'_i(x,y,z,t_n) = u_i(x,y,z,t_n) - \langle u_i(x,y,z,t_n) \rangle_{n}
\end{equation}
Invoking the ergodicity assumption and using the fluctuating velocities, the two point correlation matrix is determined as
\begin{equation}
C_{nm} = \frac{1}{N}\langle u'_i(x,y,z,t_n) u'_i(x,y,z,t_m) \rangle_{xyz}
\end{equation}
where $n,m=1,2,3,\dots,N$. The eigenvalue problem is solved to calculate the POD mode energies,
$
Ce=\lambda e
$,
which together with the correlation matrix eigenvectors $e$ can be used to calculate the time coefficients as
\begin{equation}
a^k(t_n)=\sqrt{\lambda^k N} e^k(t_n)
\end{equation}
Finally, the POD modes can be calculated as
\begin{equation}
\psi^k_i(x,y,z) = \sum_{n=1}^{N} \frac{a^k(t_n)}{\lambda^k N} u'_i(x,y,z,t_n)
\end{equation}

Some important properties regarding the POD method can be highlighted. First, the POD modes are arranged according to decreasing mode energy. This implies that a small number of POD modes can characterize the most energetic coherent structures of a turbulent flow. Secondly, POD modes can be used to reconstruct any velocity field snapshot. Finally, the resolved turbulent kinetic energy can be calculated according to
\begin{equation}
\epsilon = \langle \langle u'_i(x,y,z,t_n) u'_i(x,y,z,t_n) \rangle_{xyz} \rangle_{n} = \sum_{k=1}^{N} \lambda^k
\end{equation}

Because the ergodicity assumption, a long enough time span is necessary for the collection of the velocity field snapshots. To this end, \cite{Sirovich} proposed a nominal criterion that implies that the number of required snapshots should resolve at least $99\%$ of the turbulent kinetic energy.
\begin{equation}
\frac{\sum_{k=1}^{n} \lambda^k}{\sum_{k=1}^{\infty} \lambda^k} > 0.99
\end{equation}

\section{Results and discussions}

\subsection{Comparison to experimental results}

\begin{figure}
 \begin{center}
  \begin{adjustbox}{width=11cm}
   \begin{tikzpicture}

     \draw (0,0) rectangle (5,7.03125);     
     \draw (2.5,0) node[anchor=north] {$1.92$ m};
     \draw (2.5,-1) node[anchor=north] {a)};
     \draw (0,3.515625) node[anchor=east] {$2.7$ m};
     \draw[->,thick] (0,0) -- (5.5,0) node[anchor=north] {x};
     \draw[->,thick] (0,0) -- (0,7.53125) node[anchor=east] {z};
     
     \fill[black] (1.5625,0) rectangle (2.083,0.52083);
     \draw[<->,thick] (2.333,0) -- (2.333,0.52083);
     \draw[-] (2.133,0.52083) -- (2.533,0.52083);
     \draw (2.533,0.52083) node[anchor=west] {$0.2$ m};
     
     \draw[dashed] (0,5.2083) -- (5,5.2083);
     \draw (2.5,5.2083) node[anchor=south] {$\delta = 2.0$ m};
     
     \draw (6.5,0) rectangle (11.5,2.083);
     \draw (9,0) node[anchor=north] {$1.92$ m};
     \draw (9,-1) node[anchor=north] {b)};
     \draw (11.5,1.0415) node[anchor=west] {$0.8$ m};
     \draw[->,thick] (6.5,0) -- (12,0) node[anchor=north] {x};
     \draw[->,thick] (6.5,0) -- (6.5,2.583) node[anchor=east] {y};
     
     \fill[black] (8.0625,0.781125) rectangle (8.583,1.301875);
     \draw[<->,thick] (6.5,1.0415) -- (8.0625,1.0415);
     \draw (7.28125,1.0415) node[anchor=south] {$0.6$ m};
     \draw[<->,thick] (8.32275,0) -- (8.32275,0.781125);
     \draw (8.32275,0.3905625) node[anchor=west] {$0.3$ m};

   \end{tikzpicture}
  \end{adjustbox}
 \end{center}
  \caption{The physical dimensions of the domain used for comparison to the experimental results of Castro and Robins: a) $xz$ plane; b) $xy$ plane.}
 \label{Castro_domain}
\end{figure}
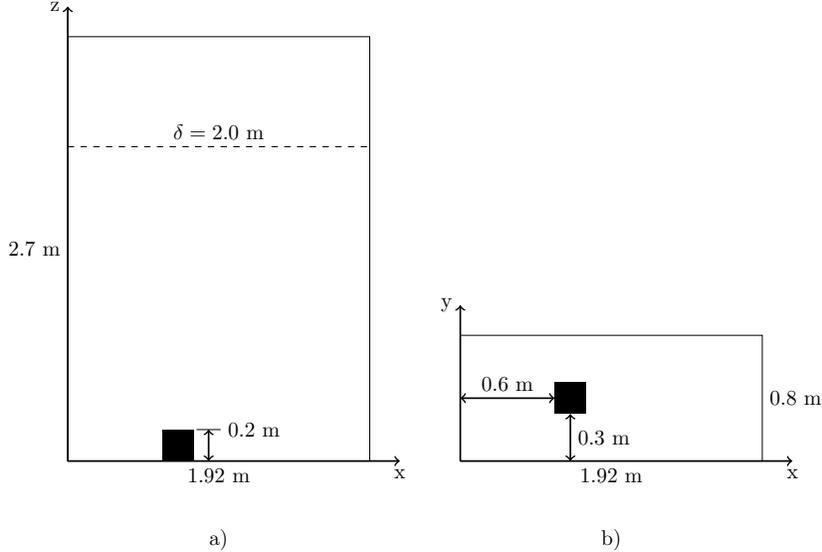

For the purpose of validating the numerical methodology, a boundary layer flow developing over a wall-mounted bluff body is considered, at a Reynolds number $Re = 3.0 \times 10^4$ based on the height of the cube and the freestream velocity. In this test case, the viscous term in the momentum equation (\ref{eq1}) is included because the viscous effects may be non-negligible. The objective of this numerical simulation is to validate the LES results with the experimental measurements taken in a wind tunnel data by \cite{Castro}. A $0.2$ m wall-mounted cube in a domain with streamwise, spanwise, and vertical dimensions of $1.92 \times 0.8 \times 2.7$ m is considered, where the height of both the domain and the cube are equal to the height of the wind tunnel test section and the experimental cube. The (streamwise, spanwise, vertical) direction names correspond to the ($x$,$y$,$z$) axes with ($u$,$v$,$w$) velocities in those directions. A representation of the physical domain is shown in figure \ref{Castro_domain}. The main domain is discretized uniformly using $192 \times 64 \times 324$ grid points in the streamwise, spanwise, and vertical directions, respectively, which gives $20 \times 16 \times 24$ grid points that are associated with the cube. The dimensions and discretization of the precursor domain are identical to the main domain. A $2.0$ m turbulent boundary layer with a freestream velocity of $2.02$ m/s is imposed in the precursor simulation using time averaged streamwise velocity, turbulence intensity, and vertical turbulent momentum flux profiles given by \cite{Castro} and the controlled forcing scheme developed by \cite{Spille}. The nondimensional time step is $\frac{\Delta t U_g}{h_{box}} = 0.005$ and the nondimensional grid spacings are $\frac{\Delta x}{h_{box}} \times \frac{\Delta y}{h_{box}} \times \frac{\Delta z}{h_{box}} = 0.05 \times 0.0625 \times 0.0417$. ($l_{box}$, $w_{box}$, $h_{box}$) are the (length, width, height) dimensions of the box and correspond to the (streamwise, spanwise, vertical) directions or ($x$,$y$,$z$) axes.

\begin{figure}
 \begin{center}
    \mbox{
        \includegraphics[width=10cm,trim=0.cm 0.2cm 0.cm 0.cm, clip=true]{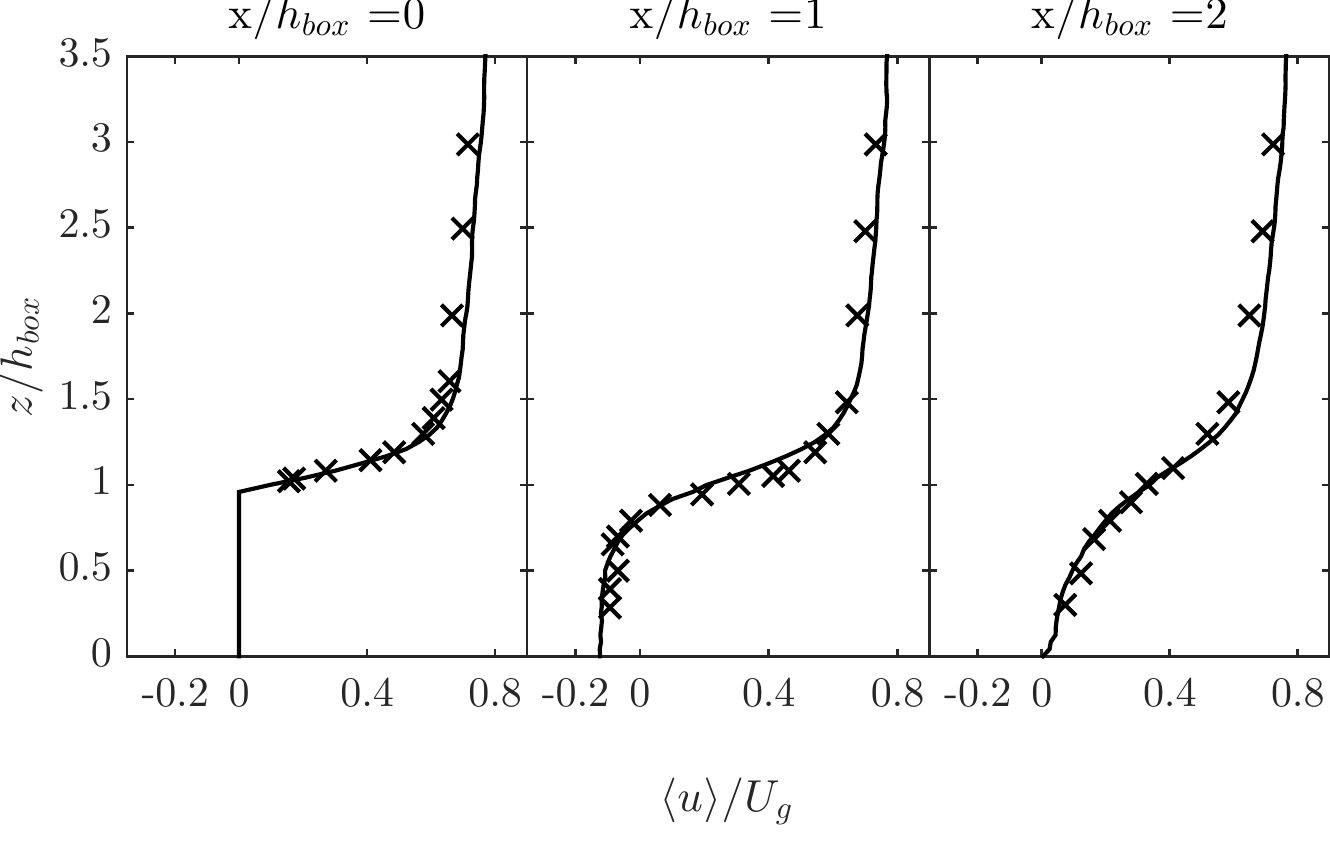}   
         }
 \end{center}
  \caption{Vertical distribution of time averaged streamwise velocity measured every box height from the rear of the box on the spanwise centerline of the box ($y/l_{box} = 0$), at conditions matching the box height Reynolds number in \cite{Castro} ($Re_{h_{box}} = 3.0 \times 10^4$):\ -----) Numerical simulation;\ X) Castro and Robins experimental data.}
  \label{uMean_Castro_z}
\end{figure}

\begin{figure}
 \begin{center} 
        \includegraphics[width=4cm,trim=0.cm -0.1cm 0.cm 0.cm, clip=true]{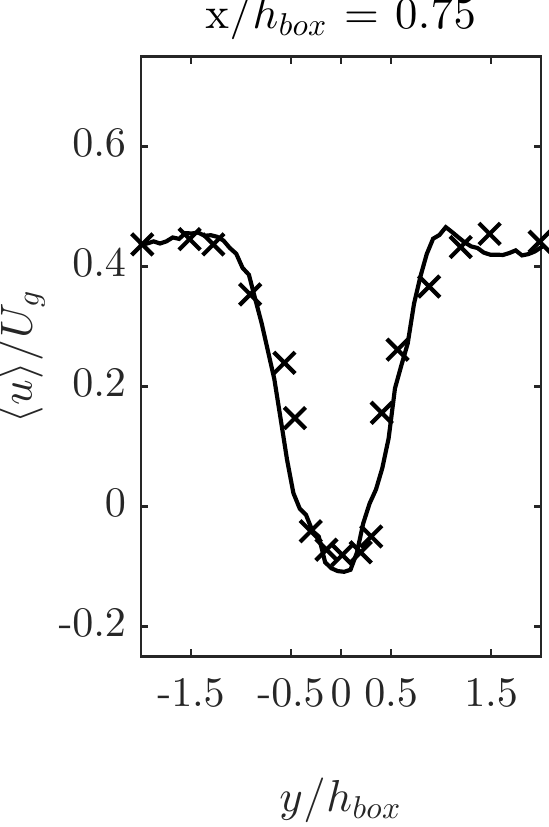}
 \end{center}
  \caption{Spanwise distribution of time averaged streamwise velocity measured at $x/h_{box} = 0.75$ and $z/h_{box} = 0.5$, at conditions matching the box height Reynolds number in \cite{Castro} ($Re_{h_{box}} = 3.0 \times 10^4$):\ -----) Numerical simulation;\ X) Castro and Robins experimental data.}
  \label{uMean_Castro_y}
\end{figure}

\begin{figure}
 \begin{center}
    \mbox{
        \includegraphics[width=10cm,trim=0.cm 0.2cm 0.cm 0.cm, clip=true]{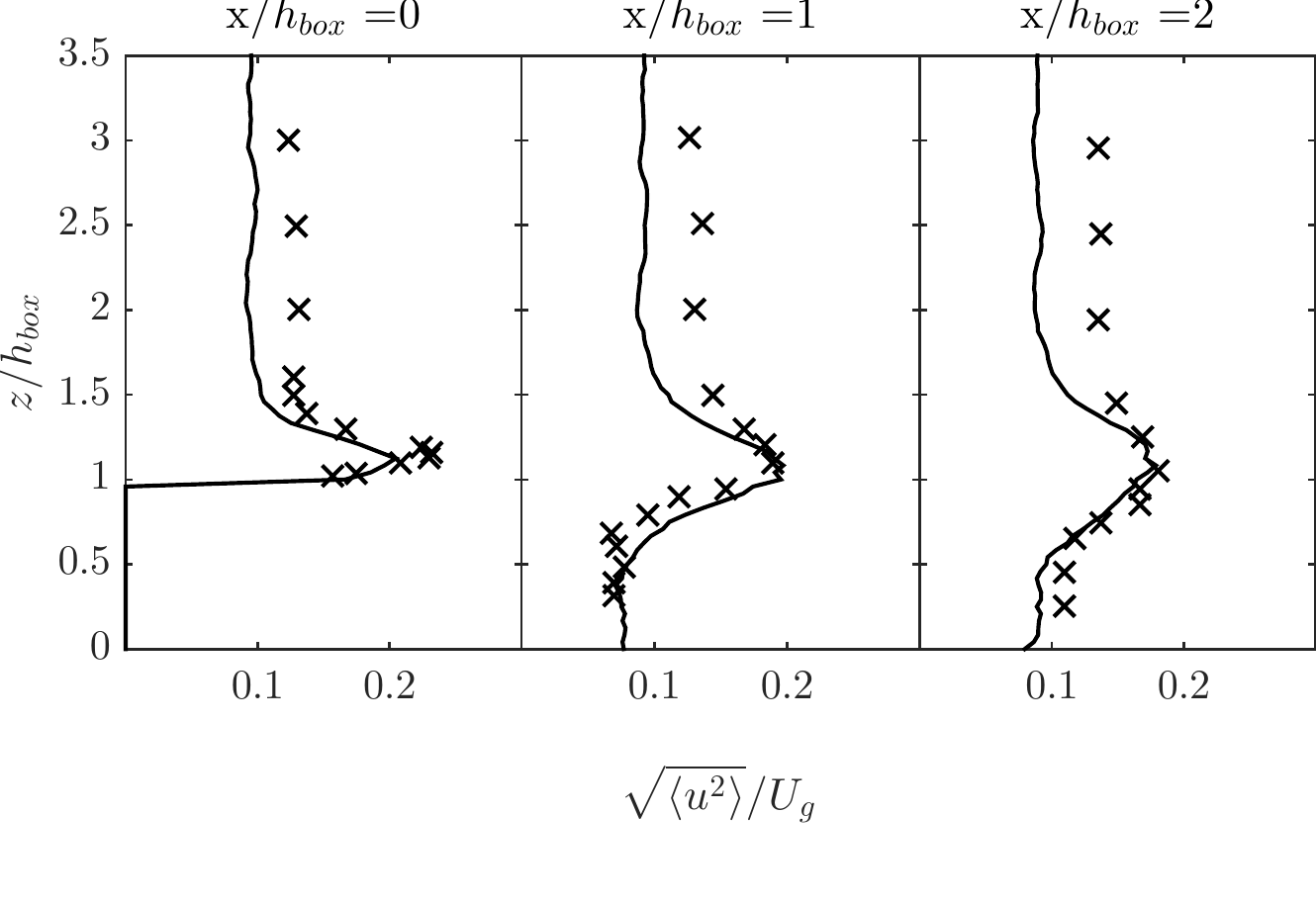}   
         }
 \end{center}
  \caption{Vertical distribution of turbulence intensity measured every box height from the rear of the box on the spanwise centerline of the box ($y/l_{box} = 0$), at conditions matching the box height Reynolds number in \cite{Castro} ($Re_{h_{box}} = 3.0 \times 10^4$):\ -----) Numerical simulation;\ X) Castro and Robins experimental data.}
  \label{varU_Castro}
\end{figure}

The time averaged streamwise velocity and turbulence intensity profiles along the vertical direction are plotted in figures \ref{uMean_Castro_z} and \ref{varU_Castro}, respectively, alongside the experimental data from Castro and Robins at three streamwise locations in the wake, $x/h_{box}= \{0,1,2\}$. Figure \ref{uMean_Castro_y} shows a spanwise profile of time averaged streamwise velocity at one streamwise location in the wake, $x/h_{box}=0.75$. The agreement was found to be good for all profiles.

\subsection{Flow around large bluff bodies}

\begin{figure}
 \begin{center}
  \begin{adjustbox}{width=10cm}
   \begin{tikzpicture}

     \draw (0,0) rectangle (12,2);     
     \draw (6,2.5) node[anchor=center] {$3770$ m};
     \draw (-1,1) node[anchor=east] {b)};
     \draw (12,1) node[anchor=west] {$628$ m};
     \draw[->,thick] (0,0) -- (12.5,0) node[anchor=north] {x};
     \draw[->,thick] (0,0) -- (0,2.5) node[anchor=east] {y};

     \fill[black] (1.5,0.875) rectangle (2.125,1.125);
     \draw[<->,thick] (0,1) -- (1.5,1);
     \draw (0.75,1) node[anchor=north] {$471$ m};
     \draw[<->,thick] (1.8125,0) -- (1.8125,0.875);
     \draw (1.8125,0.4375) node[anchor=west] {$274.5$ m};

     \draw (0,3) rectangle (12,4.909859317102744);     
     \draw (-1,4.454929658551372) node[anchor=east] {a)};
     \draw (12,4.454929658551372) node[anchor=west] {$600$ m};
     \draw[->,thick] (0,3) -- (12.5,3) node[anchor=north] {x};
     \draw[->,thick] (0,3) -- (0,5.409859317102744) node[anchor=east] {z};
     
     \fill[black] (1.5,3) rectangle (2.125,3.222816920328653);

   \end{tikzpicture}
  \end{adjustbox}
 \end{center}
  \caption{The physical dimensions of the domain used for bluff body simulations: a) $xz$ plane; b) $xy$ plane.}
 \label{Main_domain}
\end{figure}
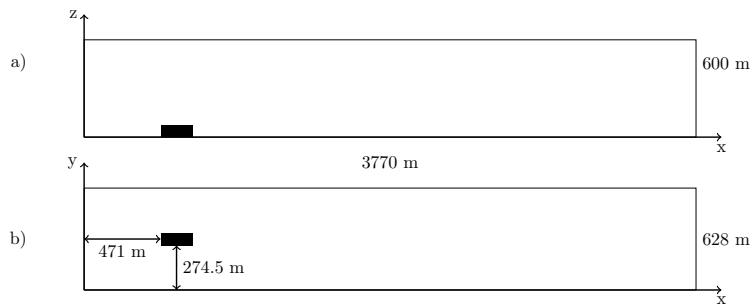

A large rectangular body with streamwise, spanwise, and vertical dimensions of $196 \times 79 \times 70$m, respectively, is considered in a stratified boundary layer flow. The body is first considered at rest, and then in motion at velocity $10$ m/s in a wind in the opposite direction of velocity of $8$ m/s. This results in a total wind velocity of $18$ m/s for the moving body case. The Reynolds numbers based on the height of the box for the stationary and moving box are both on the order of $10^7$.  Two small thermal stratifications are considered: $\Delta \theta = 0.2$ K and $\Delta \theta = 0.4$ K. The precursor and the main flow domains have the same streamwise, spanwise, and vertical dimensions of $3770 \times 628 \times 600$ m. The nondimensional time steps for the stationary and moving boxes are $\frac{\Delta t U_g}{h_{box}} = \{0.0023,0.0051\}$, respectively. Figure \ref{Main_domain} shows a model of the physical domain. Uniform grid spacing is used in the horizontal directions and also along the vertical direction because Monin-Obukhov similarity theory is used to model the flow at the wall. The nondimensional grid spacings for the $384 \times 64 \times 128$ point grid are $\frac{\Delta x}{h_{box}} \times \frac{\Delta y}{h_{box}} \times \frac{\Delta z}{h_{box}} = 0.14 \times 0.14 \times 0.06$. The difference in potential temperature between the ground, $\theta_s$, and the top of the ABL, $\theta_{t}$, is used to set the thermal stratification, $\Delta \theta = \theta_{t} - \theta_{s}$.

\subsubsection{Contour plots and vertical profiles}

\begin{figure}
 \begin{center}
    \mbox{
     a) \includegraphics[width=10cm]{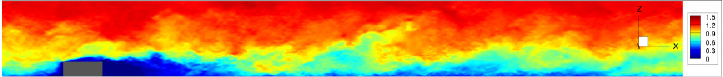}
         }
    \mbox{
     b) \includegraphics[width=10cm]{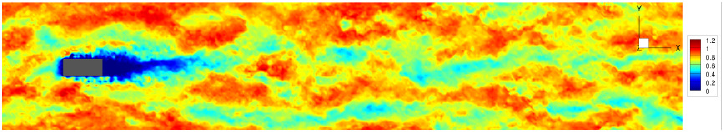}
         } \\
          \vspace{3mm} 
    \mbox{
      c) \includegraphics[width=10cm]{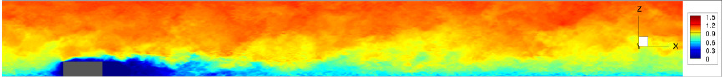}
         }
    \mbox{
      d) \includegraphics[width=10cm]{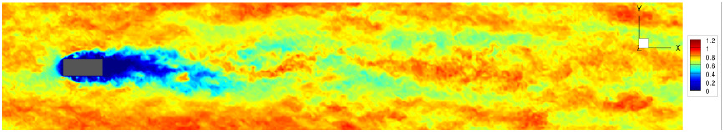}
         }

 \end{center}
  \caption{Instantaneous streamwise velocity normalized by the wind velocity contours on vertical and horizontal planes sectioning the box through the center for a thermal stratification of $\Delta\theta = 0.2$ K): a) $xz$ plane and with $U_g=8$ m/s; b) $xy$ plane and $U_g=8$ m/s; c) $xz$ plane and $U_g=18$ m/s; d) $xy$ plane and $U_g=18$ m/s.}
  \label{instant_u}
\end{figure}

\begin{figure}
 \begin{center}
    \mbox{
      a) \includegraphics[width=10cm]{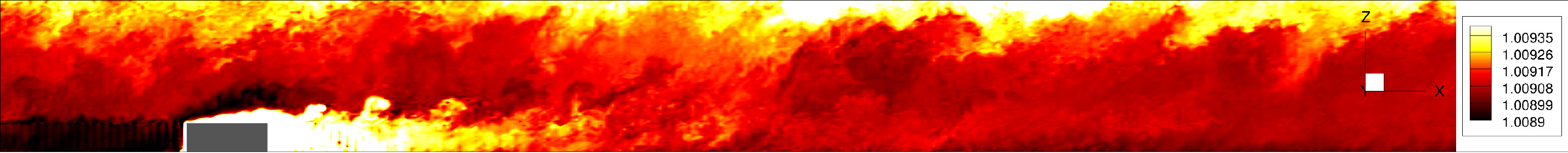}   
         }
    \mbox{
      b) \includegraphics[width=10cm]{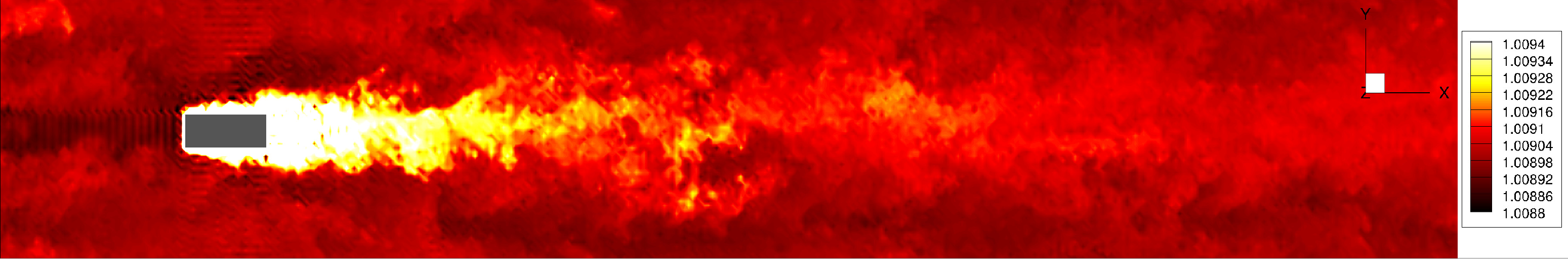}
         }\\
         \vspace{3mm} 
    \mbox{
      c) \includegraphics[width=10cm]{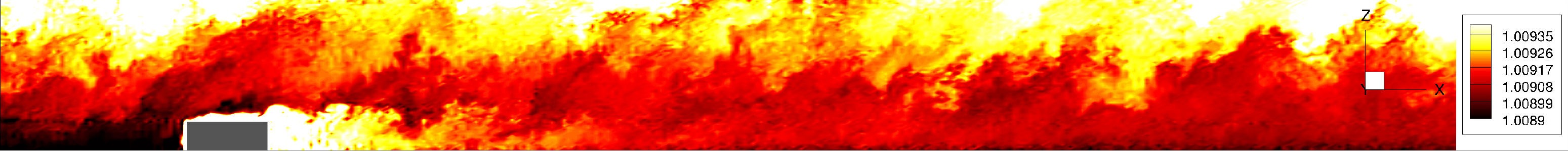} 
         }
    \mbox{
      d) \includegraphics[width=10cm]{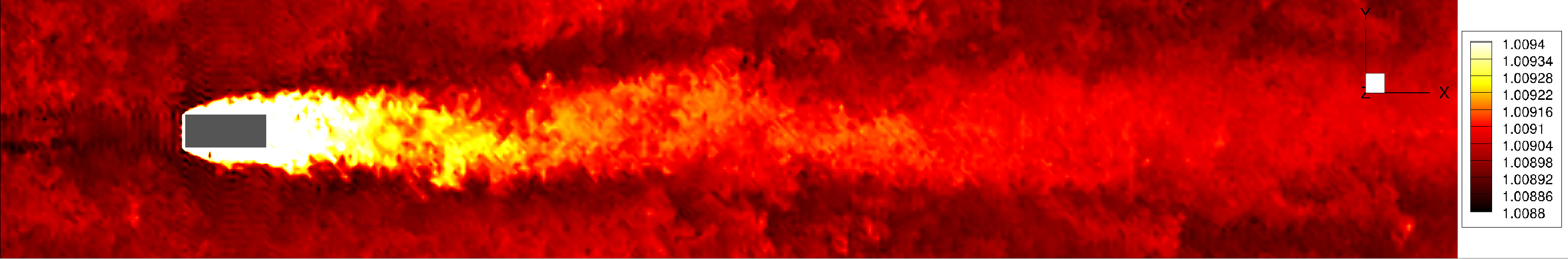}
         }

 \end{center}
  \caption{Potential temperature normalized by the ground temperature contours on vertical and horizontal planes sectioning the box through the center for a thermal stratification of $\Delta\theta = 0.2$ K): a) $xz$ plane and with $U_g=8$ m/s; b) $xy$ plane and $U_g=8$ m/s; c) $xz$ plane and $U_g=18$ m/s; d) $xy$ plane and $U_g=18$ m/s.}
  \label{instant_theta}
\end{figure}

Several contours of instantaneous streamwise velocity and potential temperature  are presented in figures \ref{instant_u} and \ref{instant_theta}. In particular, figures \ref{instant_u}(a) and \ref{instant_u}(b) illustrate the streamwise component of velocity using contours through $xz$ and $xy$ planes, respectively, sectioning the box through the center. It shows that the wake is more intense in the proximity to the box, about two to three box lengths in the downstream. Since the thermal stratification is small, large turbulent structures, in the same order of magnitude as the size of the box, are moving with the flow in the downstream direction (they interact with the box and and the wake). In figures \ref{instant_u}(c) and \ref{instant_u}(d), corresponding to a wind velocity of $18$ m/s, one can notice that the wake becomes more intense and persists for a longer distance. They also show that the length of the wake along the spanwise direction is larger than the length of the wake generated by the box moving against a $8$ m/s wind. In figures \ref{instant_theta}(a) and \ref{instant_theta}(b), contour plots of the potential temperature through an $xz$ plane and an $xy$ plane, respectively, sectioning the box in the center, are depicted. The thermal wake seems to be much longer than the momentum wake (light red and yellow are associated with larger temperatures). Similar contour plots of potential temperature are shown in figures \ref{instant_theta}(c) and \ref{instant_theta}(d) for the wind velocity $18$ m/s. One can notice that the thermal wake is more compact in the case of the larger incoming velocity.


The vertical profiles shown in the next figures \ref{varU_02}-\ref{varWTheta_10} represent time averaged flow data that was plotted in the downstream of the box at every two box lengths from the center of the box. In all figures, the vertical spatial coordinate is scaled by the height of the box. The streamwise time averaged turbulence intensity is revealed in figures \ref{varU_02}  and \ref{varU_10} for various cases. In figure \ref{varU_02}, one can notice that by increasing the wind velocity, the turbulence intensity in the proximity to the box is enhanced starting from the bottom boundary to approximately one and a half box heights. As the streamwise distance is increased, the turbulence intensity difference between the two cases diminishes. The increase in the thermal stratification slightly impacts the vertical distribution of the turbulence intensity as shown in figure \ref{varU_10}.  Only very small differences can be noticed in the vicinity of the box (this is because the thermal stratification is very small).

\begin{figure}
 \begin{center}
        \mbox{
        \includegraphics[width=10cm,trim=0.cm 0.2cm 0.cm 0.cm, clip=true]{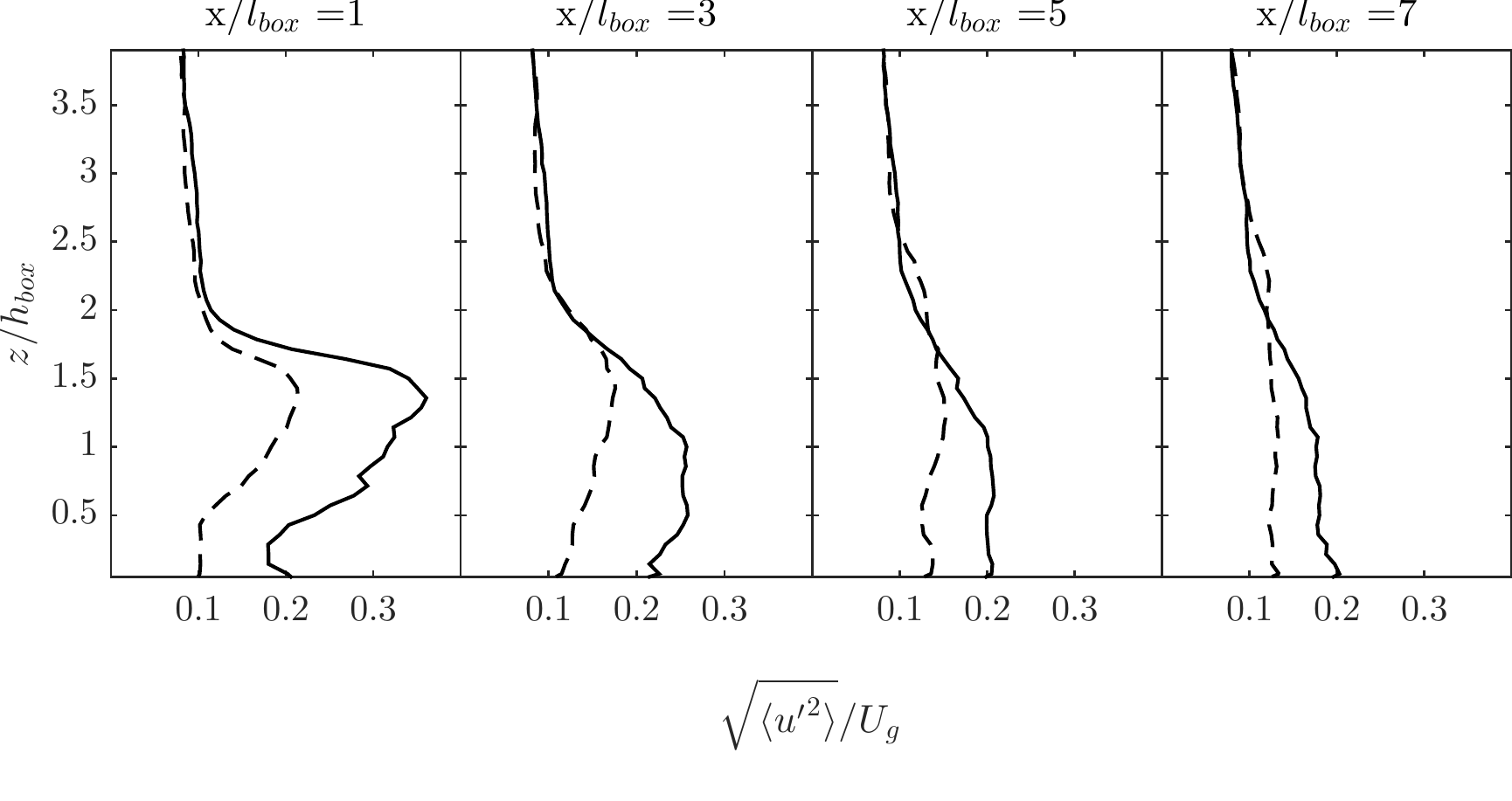}
         }
 \end{center}
  \caption{Vertical distribution of turbulence intensity measured every two box lengths from the rear of the body ($\Delta\theta = 0.2$ K):\ -\ -\ -) $U_g=8$ m/s;\ -----) $U_g=18$ m/s.}
  \label{varU_02}
\end{figure}

\begin{figure}
 \begin{center}
        \mbox{
        \includegraphics[width=10cm,trim=0.cm 0.2cm 0.cm 0.cm, clip=true]{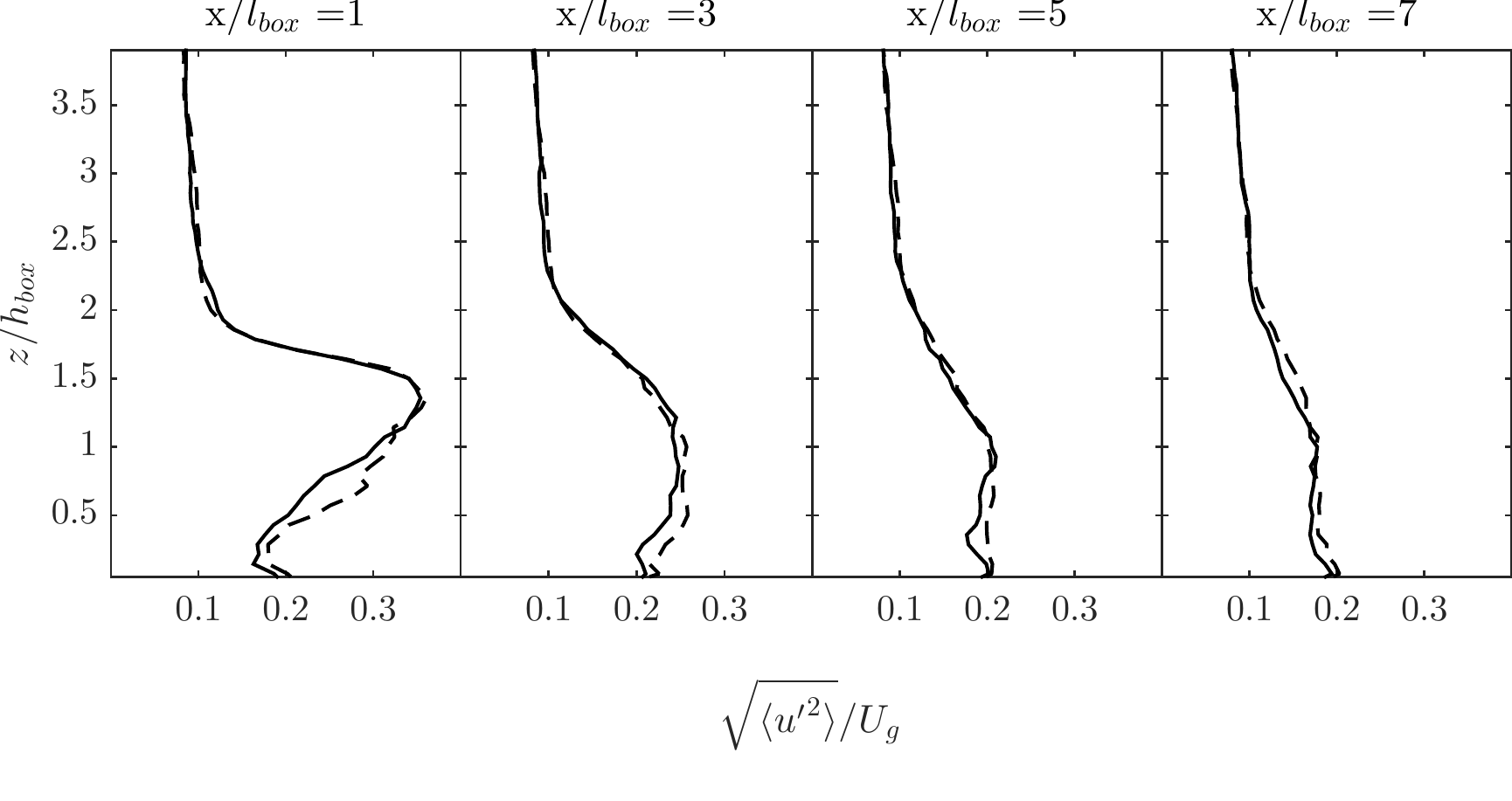}
         }
 \end{center}
  \caption{Vertical distribution of turbulence intensity measured every two box lengths from the rear of the body ($U_g=18$ m/s):\ -\ -\ -) $\Delta\theta = 0.2$ K;\ -----) $\Delta\theta = 0.4$ K.}
  \label{varU_10}
\end{figure}

Figures \ref{varUW_02} and \ref{varUW_10} present profiles of the vertical turbulent momentum flux at four locations downstream of the body. In figure \ref{varUW_02}, the thermal stratification is kept constant at $0.2$, and averaged profiles from simulation cases with wind velocities $8$ m/s and $18$ m/s are compared to each other. One can notice that by increasing the freestream velocity the vertical momentum flux magnitude decreases, with a maximum attained in the vicinity of the top of the box. A small increase in the momentum flux can be noted in the proximity to the ground. There are clear differences between the profiles for the two wind velocities beyond seven box lengths downstream from the box. In figure \ref{varUW_10}, the wind velocity is constant at $18$ m/s and averaged profiles for the thermal stratifications of $0.2$ and $0.4$ are compared to each other. Only very small differences are noted in the vicinity of the box.

\begin{figure}
 \begin{center}
        \mbox{
        \includegraphics[width=10cm,trim=0.cm 0.2cm 0.cm 0.cm, clip=true]{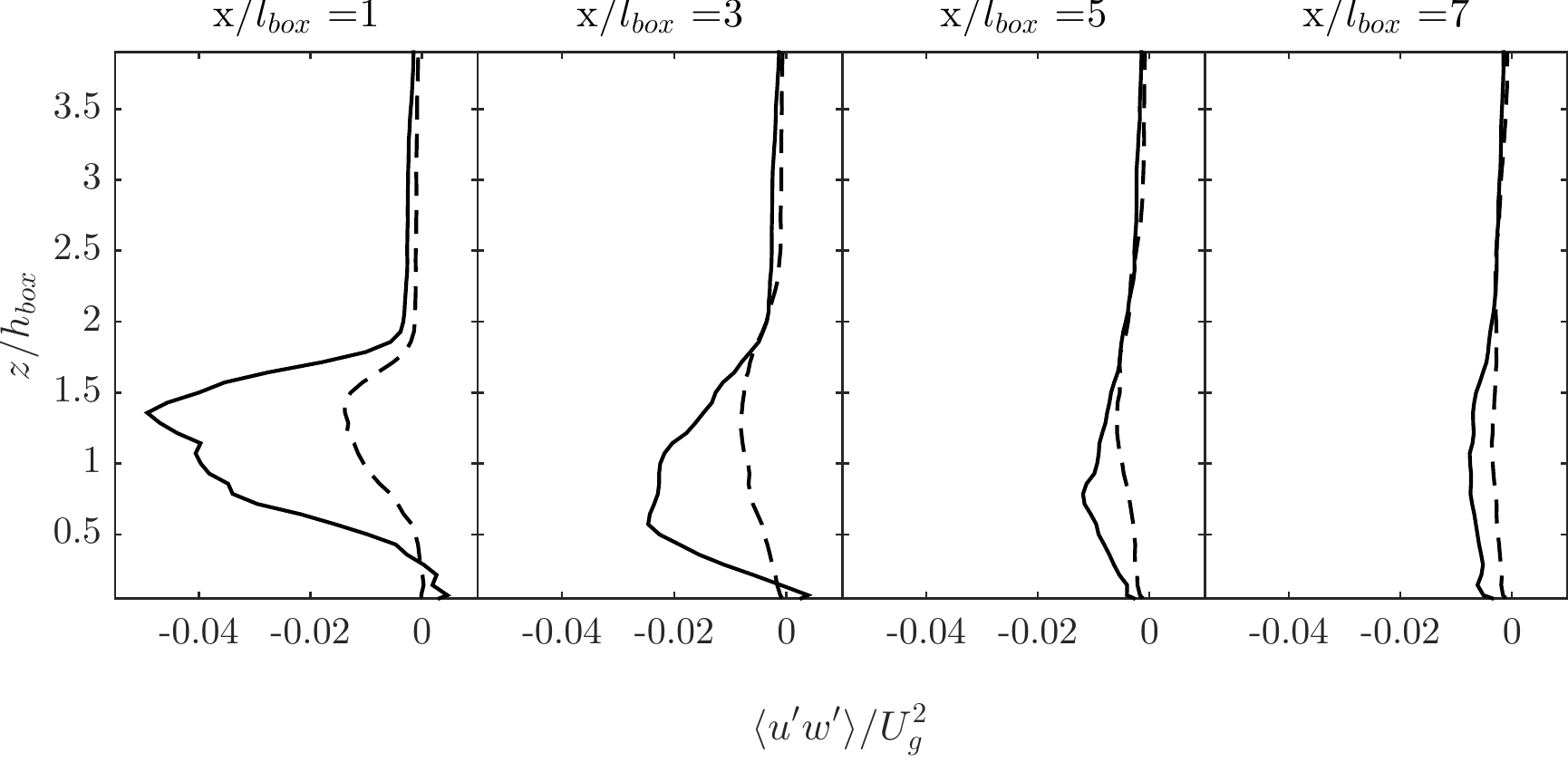}
         }
 \end{center}
  \caption{Vertical distribution of vertical turbulent momentum flux measured every two box lengths from the rear of the body ($\Delta\theta = 0.2$ K):\ -\ -\ -) $U_g=8$ m/s;\ -----) $U_g=18$ m/s.}
  \label{varUW_02}
\end{figure}

\begin{figure}
 \begin{center}
        \mbox{
        \includegraphics[width=10cm,trim=0.cm 0.2cm 0.cm 0.cm, clip=true]{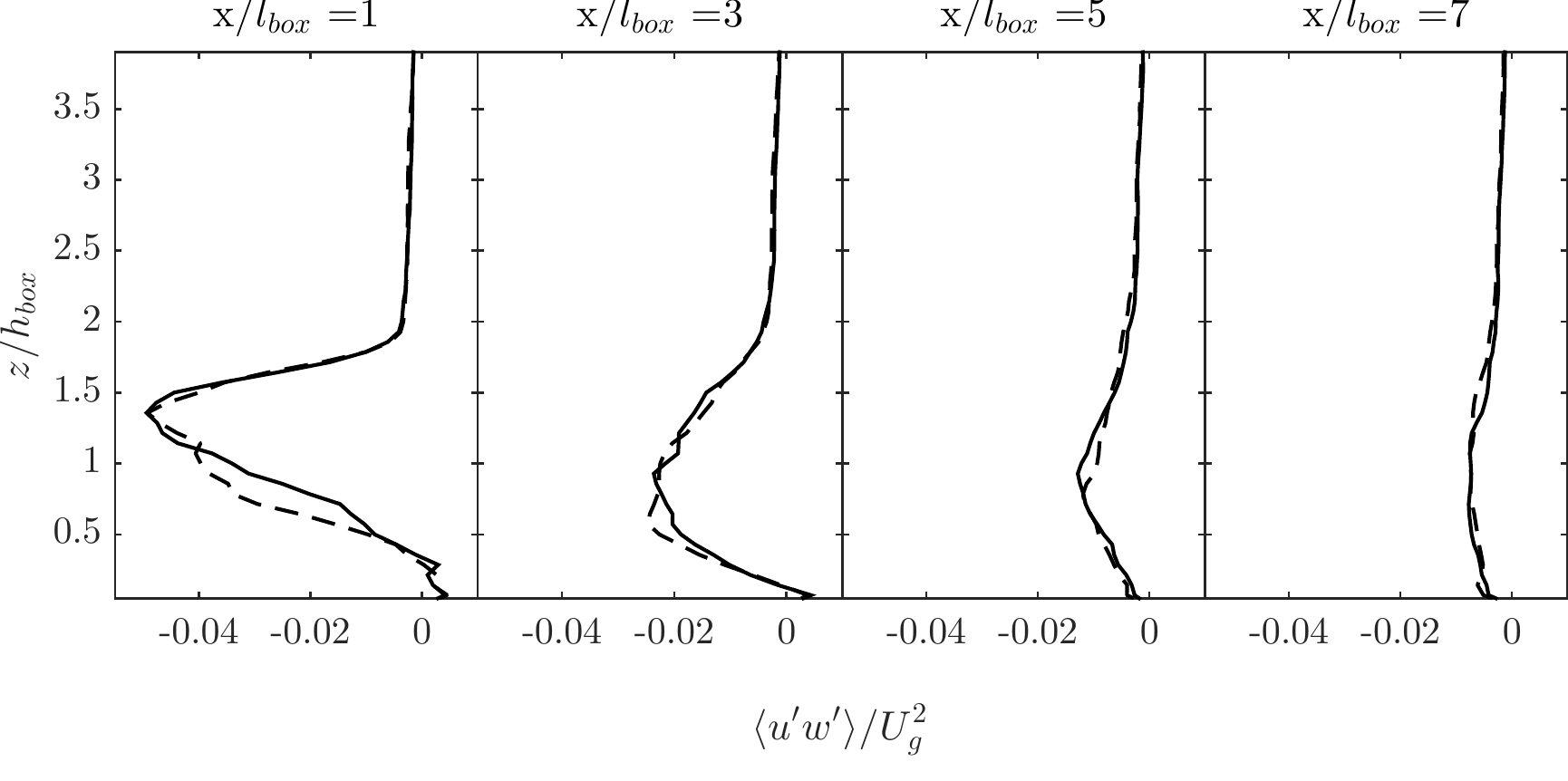}
         }
 \end{center}
  \caption{Vertical distribution of vertical turbulent momentum flux measured every two box lengths from the rear of the body ($U_g=18$ m/s):\ -\ -\ -) $\Delta\theta = 0.2$ K;\ -----) $\Delta\theta = 0.4$ K.}
  \label{varUW_10}
\end{figure}

In the next two figures, comparisons in terms of the vertical turbulent heat flux are presented. Profiles of vertical turbulent heat flux are shown in figure \ref{varWTheta_02} for a stratification of $\Delta\theta = 0.2$ K in order to compare the both wind velocity cases. The magnitude of the vertical turbulent heat flux behind the box is larger in the case of the larger freestream velocity, up to approximately $x/l_{box}=3$; at this location, there is a decrease in the turbulent flux above the box. By seven box lengths in the downstream, there is no significant differences between the two profiles. Figure \ref{varWTheta_10} shows that there are no noticeable differences in the heat flux profiles between the two thermal stratifications.

\begin{figure}
 \begin{center}
        \mbox{
        \includegraphics[width=10cm,trim=0.cm 0.2cm 0.cm 0.cm, clip=true]{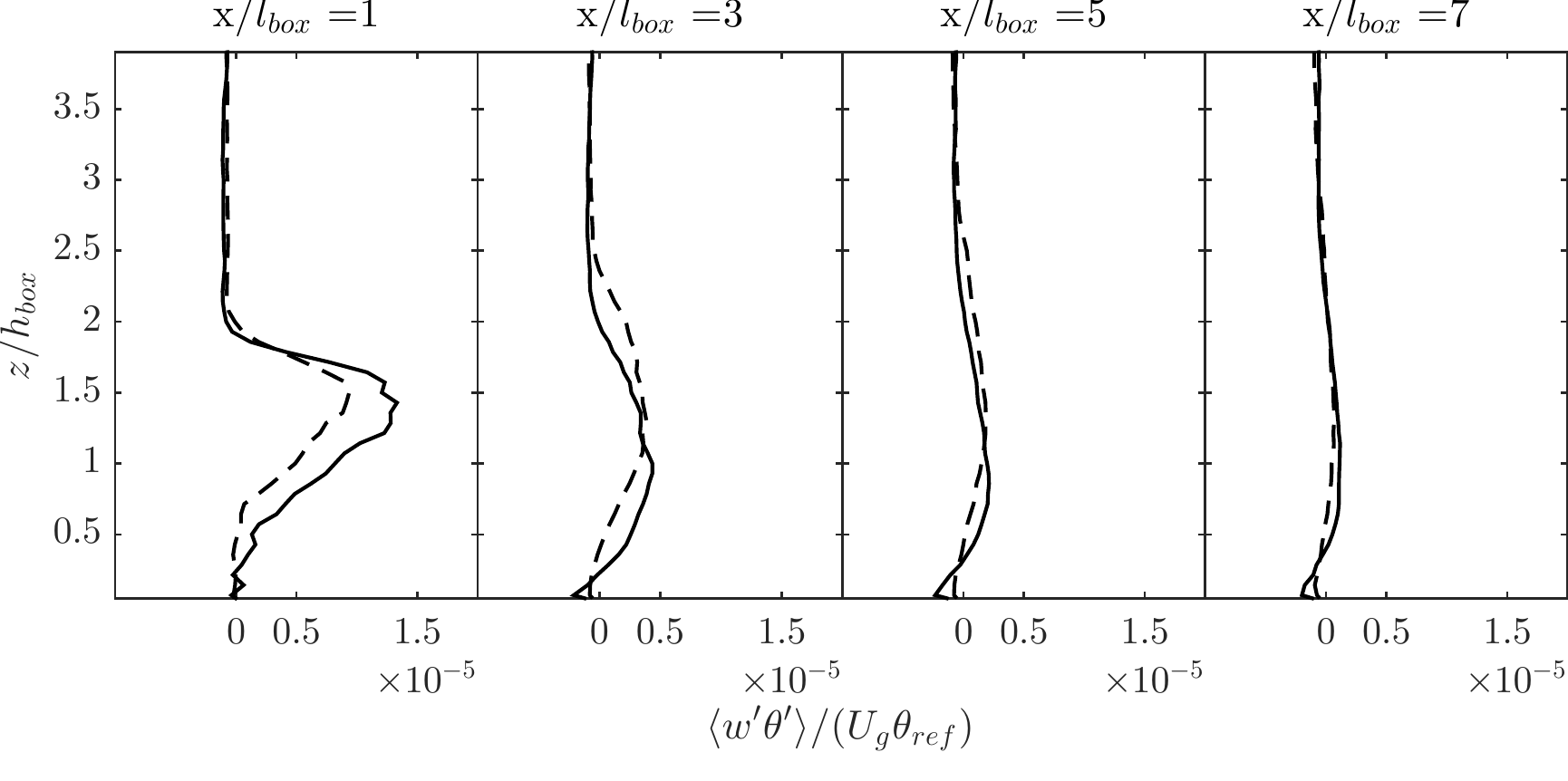}
         }
 \end{center}
  \caption{Vertical distribution of vertical turbulent heat flux measured every two box lengths from the rear of the body ($\Delta\theta = 0.2$ K):\ -\ -\ -) $U_g=8$ m/s;\ -----) $U_g=18$ m/s.}
  \label{varWTheta_02}
\end{figure}

\begin{figure}
 \begin{center}
        \mbox{
        \includegraphics[width=10cm,trim=0.cm 0.2cm 0.cm 0.cm, clip=true]{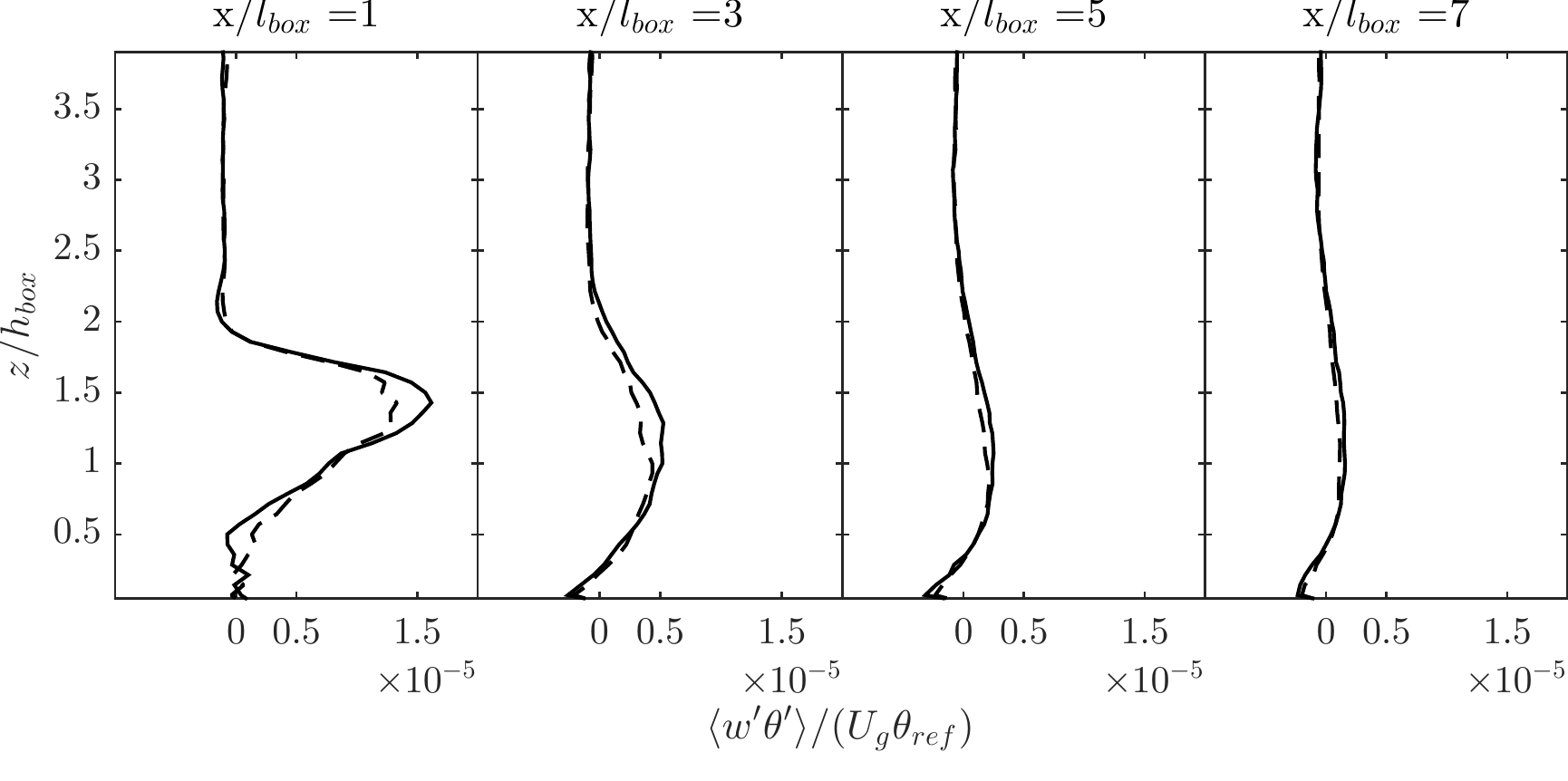}
         }
 \end{center}
  \caption{Vertical distribution of vertical turbulent heat flux measured every two box lengths from the rear of the body ($U_g=18$ m/s):\ -\ -\ -) $\Delta\theta = 0.2$ K;\ -----) $\Delta\theta = 0.4$ K.}
  \label{varWTheta_10}
\end{figure}

\subsubsection{POD results}
\begin{figure}
 \begin{center}
        \includegraphics[width=10cm]{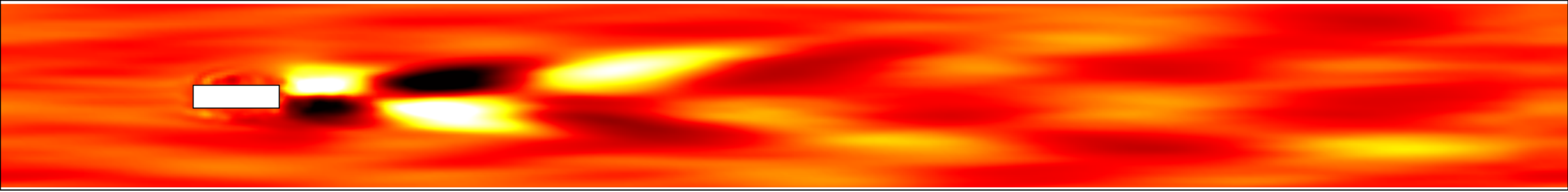}   \\
         a)
 \end{center}
 \begin{center}
        \includegraphics[width=10cm]{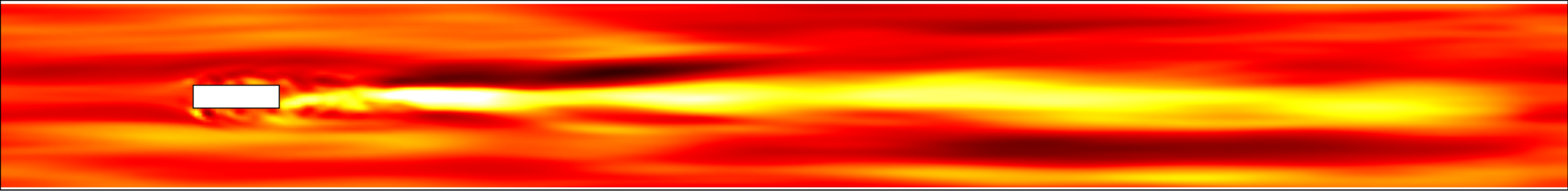}   \\
         b)
 \end{center}
 \begin{center}
        \includegraphics[width=10cm]{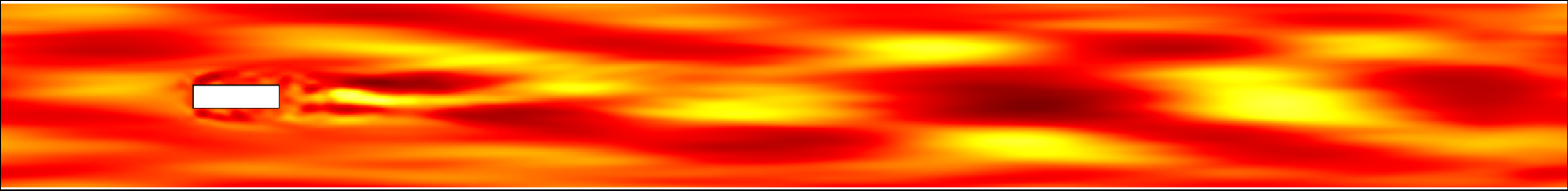}   \\
         c)
 \end{center}
 \begin{center} 
        \includegraphics[width=3.9cm]{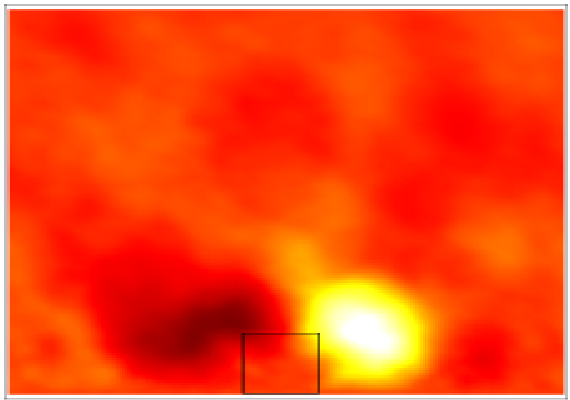}   
        \includegraphics[width=3.9cm]{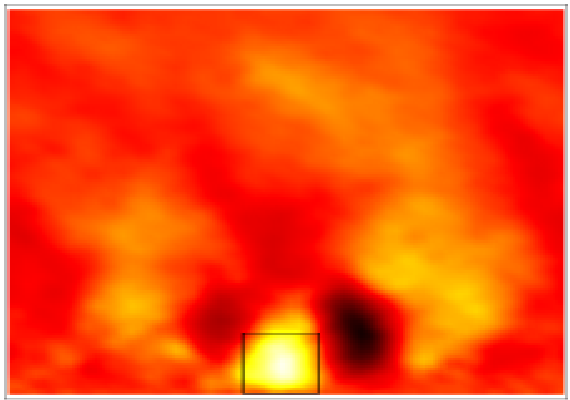}   \\
         d)  \hspace{33mm} e)
 \end{center}
  \caption{Contours of different types of typical streamwise velocity POD modes from various cases: a) helical mode; b) symmetric mode; c) combined helical-symmetric mode; d) cross-section through the wake revealing a helical mode; e) cross-section through the wake revealing a symmetric mode.}
  \label{mode_types}
\end{figure}

The snapshot POD analysis was performed utilizing the same tool used in \cite{VerHulst}. The three-dimensional velocity field was saved every $0.2$ seconds for a total of 8000 snapshots. When first inspecting contours of the streamwise velocity POD modes, three main mode types were identified in the region of the wake: helical, symmetric, and combined helical-symmetric. Figures \ref{mode_types}(a), \ref{mode_types}(b), and \ref{mode_types}(c) show examples of the three types of POD modes, while figures \ref{mode_types}(d) and \ref{mode_types}(e) show the helical and symmetric modes, respectively, using contours cross-sectioning the wake. 

\begin{figure}
 \begin{center}
        \includegraphics[width=5cm]{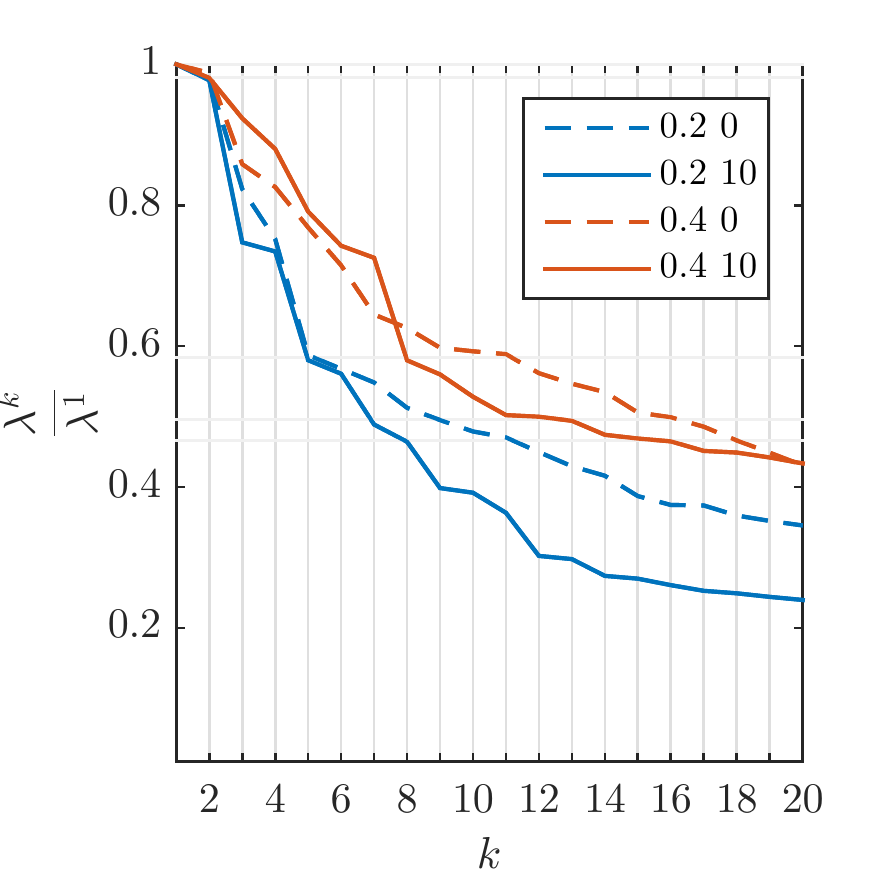}
	\includegraphics[width=5cm]{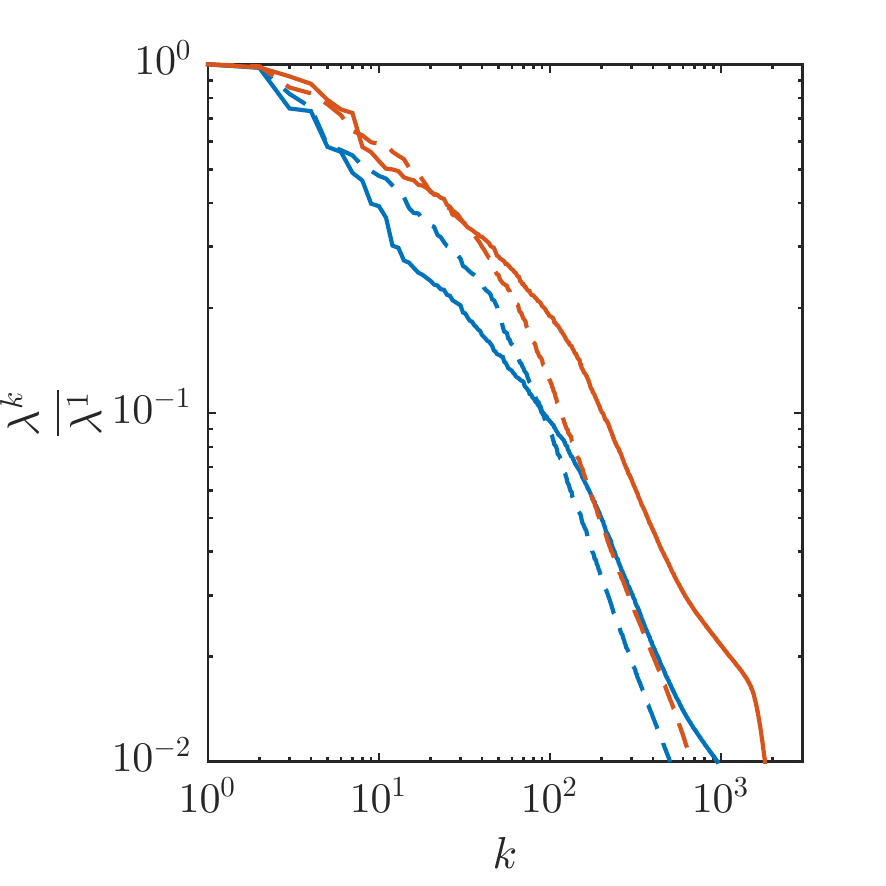}
	 \\
      \hspace{6mm} a) \hspace{38mm} b)
 \end{center}
  \caption{POD mode energies normalized with the first mode energy: a) the first twenty modes; b) all the modes at least $1\%$ as energetic as the first mode; blue) $\Delta\theta = 0.2$ K; red) $\Delta\theta = 0.4$ K; -\ -\ -) $U_g=8$ m/s; -----) $U_g=18$ m/s.}
  \label{mode_energy}
\end{figure}

In figure \ref{mode_energy}(a), the energy levels of first twenty POD modes normalized by the highest energy containing mode are shown for all cases. The relative mode energy levels decrease rapidly over the first twenty POD modes and by the twentieth mode, the corresponding energy is less than half of the energy contained in the strongest mode. So, the behavior of these first modes is important in representing the most energetic flow structures. Figure \ref{mode_energy}(b) plots the relative energy levels of all the modes that contain at least one percent of energy of the strongest POD mode on a log-log scale. Table \ref{mode_TKE} highlights the mode number at the one percent threshold. When comparing at constant thermal stratification, the initial modes of the lower wind velocity cases contain more of the relative energy than the higher wind velocity cases until modes $k=90$ and $k=35$ for the $\Delta \theta = 0.2$ K and $\Delta \theta = 0.4$ K thermal stratification levels, respectively. After those crossover modes, the modes of the higher wind velocity cases are at higher relative energy levels. Increasing the wind velocity spreads the turbulent kinetic energy over a larger number of POD modes. This can be seen in the mode number of the one percent threshold; there are $1.9$ times, for $\Delta \theta = 0.2$ K, and $2.7$ times, for $\Delta \theta = 0.4$ K, more modes with energy levels above the one percent threshold. Holding wind velocity constant, increasing the thermal stratification raised the relative energy levels of all the modes. 

\begin{figure}
 \begin{center}
        \includegraphics[width=10cm]{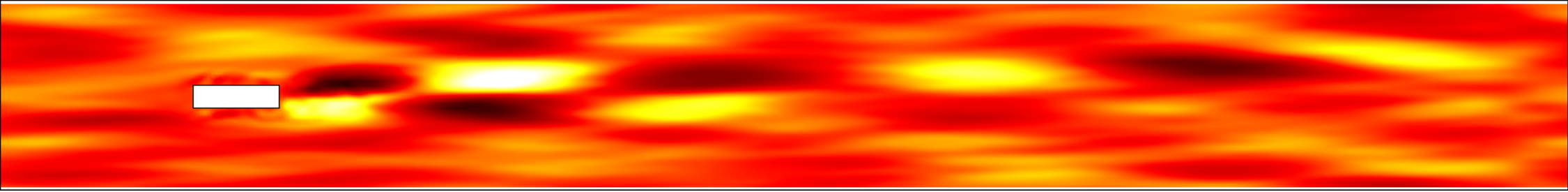}   \\
         a)
 \end{center} 
 \begin{center}
        \includegraphics[width=10cm]{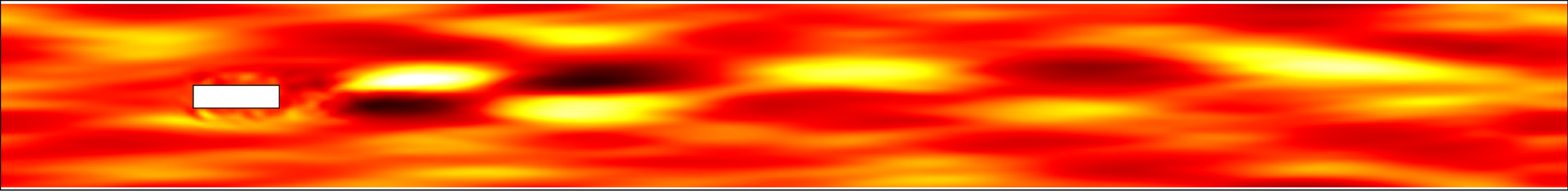}   \\
         b)
 \end{center}
 \begin{center}
        \includegraphics[width=10cm]{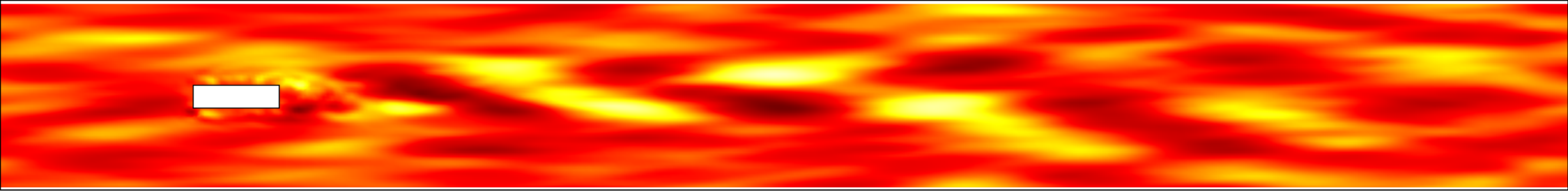}   \\
         c)
 \end{center}
 \begin{center}
        \includegraphics[width=10cm]{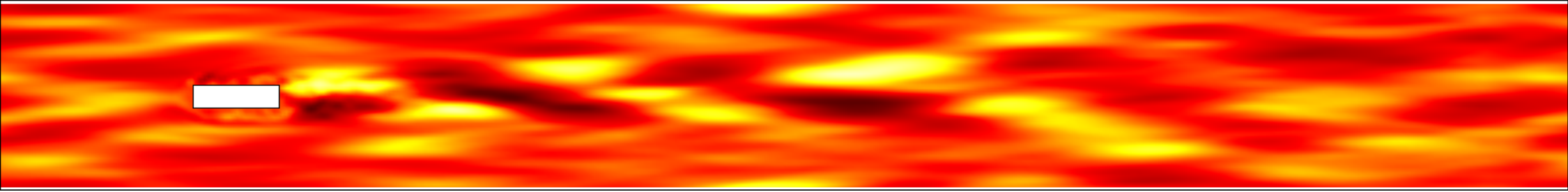}   \\
         d)
 \end{center}
  \caption{Streamwise velocity POD mode contours on a horizontal plane sectioning the box through the center, highlighting the paired helical modes at different wavenumbers.}
  \label{mode_pair}
\end{figure}

Notice that in figure \ref{mode_energy}(a), certain consecutive modes occur at nearly equal relative energy levels: $k=1$ and $k=2$ for $\Delta\theta = 0.4$ K and $U_g=8$ m/s and $k=9$ and $k=10$ for $\Delta\theta = 0.4$ K and $U_g=8$ m/s. These consecutive modes are a pair of similar helical modes differentiated by a wavelength shifting. In order to preserve the orthogonality of the two modes, the wavelength of one of the modes is shifted by a quarter of the  wavelength of the other mode. Figures \ref{mode_pair}(a) and \ref{mode_pair}(b) show two examples, at different wavenumbers, of the helical mode pairing. For each pair, the wavenumbers are the same, but the modes are only shifted by a quarter wavelength. By looking at figure \ref{mode_energy}(a), the number of paired helical modes at the highest relative energy levels increases with increasing wind velocity and thermal stratification. Through this analysis of the significant energy contributing POD modes, the helical mode was found to be dominant.

\begin{table}[htpb]
 \begin{center}
  \begin{tabular}{| c || c | c |} \hline
                               & $U_g = 8$ m/s     & $U_g = 18$ m/s \\	\hline \hline
       $\Delta\theta = 0.2$ K  & $504$             & $953$	      \\	\hline
       $\Delta\theta = 0.4$ K  & $673$             & $1807$  	      \\
\hline
  \end{tabular}
    \caption{POD mode number ($k$) where $\frac{\lambda^k}{\lambda^1} < 0.01$.}
  \label{mode_TKE}
 \end{center}
\end{table}

\begin{figure}
 \begin{center}
    \begin{adjustbox}{width=14cm}
        	      k=1\hspace{2.5mm} \includegraphics[width=6.3cm,trim=0.cm 0.cm 1.cm 0.cm, clip=true]{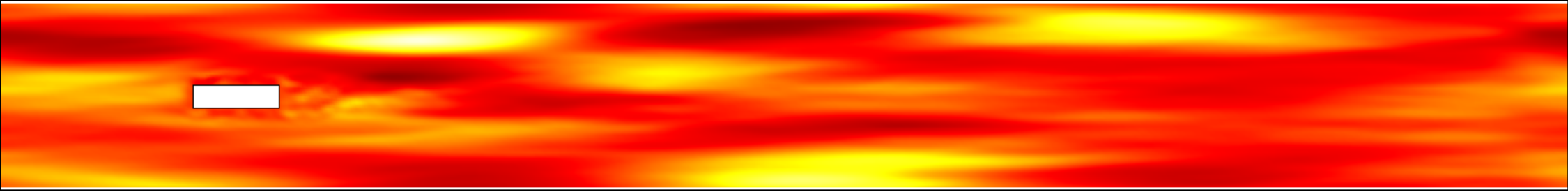}
        \hspace{10.mm}k=1\hspace{2.5mm}	\includegraphics[width=6.3cm,trim=0.cm 0.cm 1.cm 0.cm, clip=true]{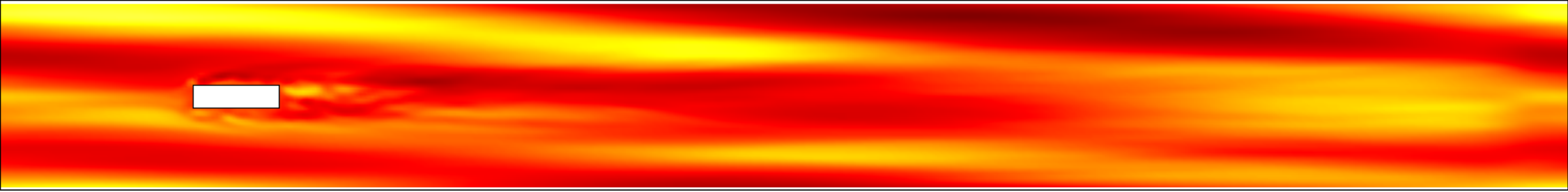}
    \end{adjustbox}
    \begin{adjustbox}{width=14cm}
        	      k=2\hspace{2.5mm} \includegraphics[width=6.3cm,trim=0.cm 0.cm 1.cm 0.cm, clip=true]{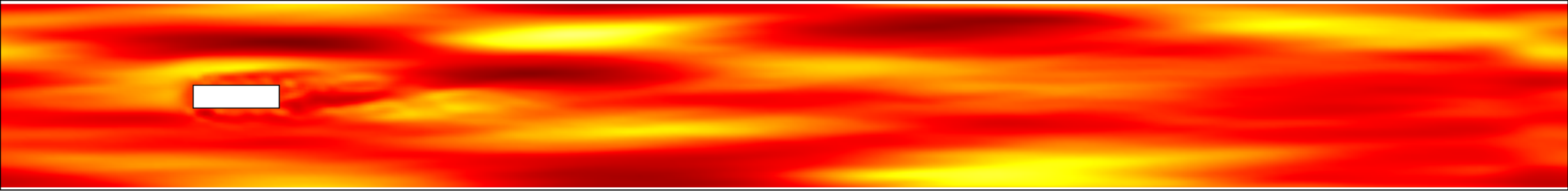}
        \hspace{10.mm}k=2\hspace{2.5mm}	\includegraphics[width=6.3cm,trim=0.cm 0.cm 1.cm 0.cm, clip=true]{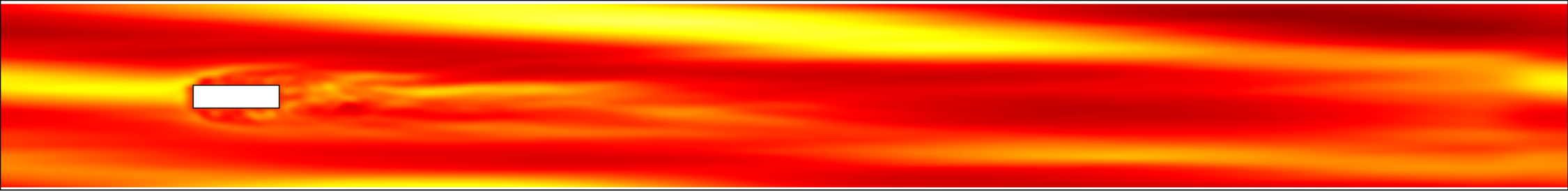}
    \end{adjustbox}
    \begin{adjustbox}{width=14cm}
        	      k=5\hspace{2.5mm} \includegraphics[width=6.3cm,trim=0.cm 0.cm 1.cm 0.cm, clip=true]{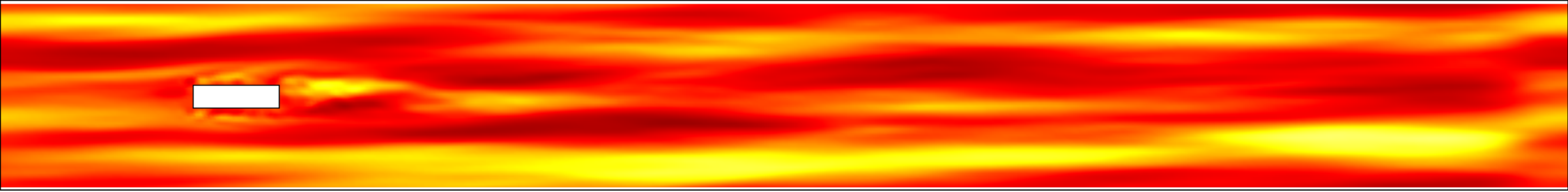}
        \hspace{10.mm}k=5\hspace{2.5mm}	\includegraphics[width=6.3cm,trim=0.cm 0.cm 1.cm 0.cm, clip=true]{u_box_02_10_xy_mode_05}
    \end{adjustbox}
    \begin{adjustbox}{width=14cm}
        	      k=9\hspace{2.5mm} \includegraphics[width=6.3cm,trim=0.cm 0.cm 1.cm 0.cm, clip=true]{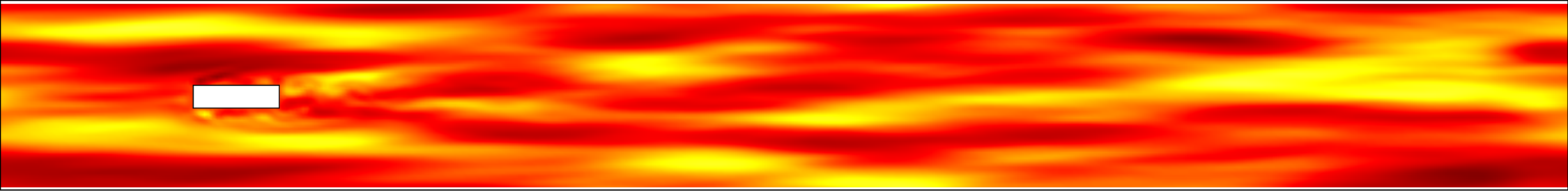}
        \hspace{10.mm}k=7\hspace{2.5mm}	\includegraphics[width=6.3cm,trim=0.cm 0.cm 1.cm 0.cm, clip=true]{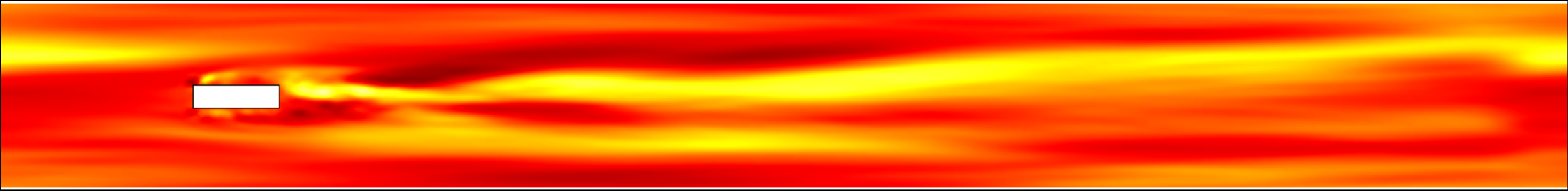}
    \end{adjustbox}
    \begin{adjustbox}{width=14cm}
        	      k=11\hspace{1.5mm}\includegraphics[width=6.3cm,trim=0.cm 0.cm 1.cm 0.cm, clip=true]{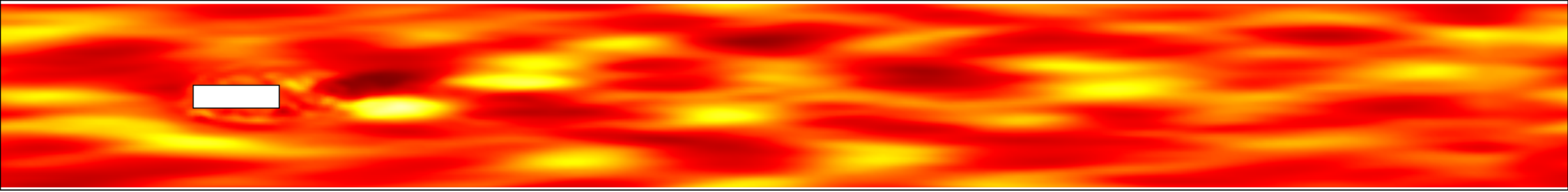}
        \hspace{10.mm}k=8\hspace{2.5mm}	\includegraphics[width=6.3cm,trim=0.cm 0.cm 1.cm 0.cm, clip=true]{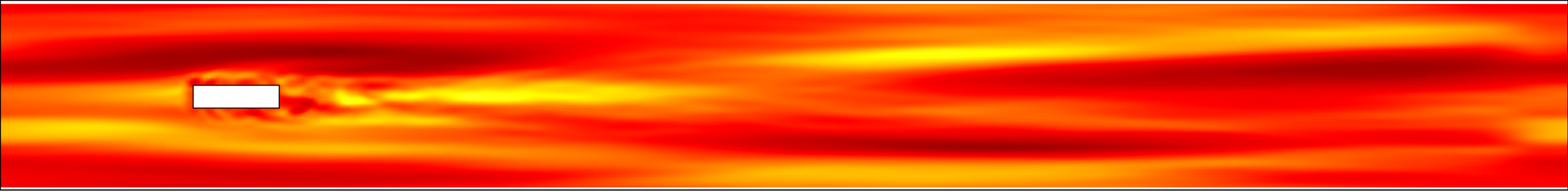}
    \end{adjustbox} \\
      a) \hspace{40mm}  b) \\ \vspace{2.5mm}   
   
    \begin{adjustbox}{width=14cm}
        	      k=1\hspace{2.5mm} \includegraphics[width=6.3cm,trim=0.cm 0.cm 1.cm 0.cm, clip=true]{u_box_04_00_xy_mode_01}
        \hspace{10.mm}k=1\hspace{2.5mm} \includegraphics[width=6.3cm,trim=0.cm 0.cm 1.cm 0.cm, clip=true]{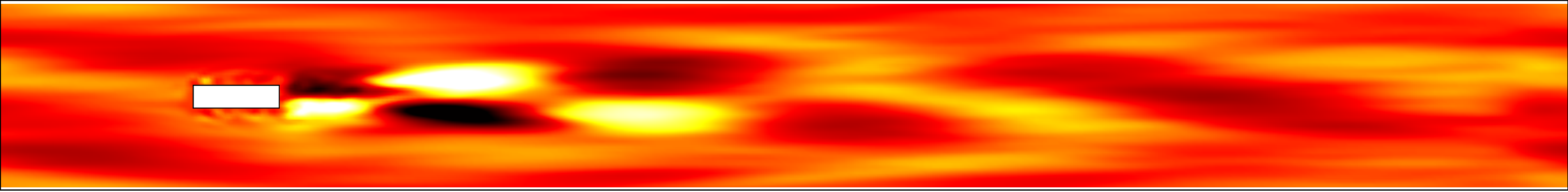}
    \end{adjustbox}
    \begin{adjustbox}{width=14cm}
        	      k=2\hspace{2.5mm} \includegraphics[width=6.3cm,trim=0.cm 0.cm 1.cm 0.cm, clip=true]{u_box_04_00_xy_mode_02}
        \hspace{10.mm}k=2\hspace{2.5mm}	\includegraphics[width=6.3cm,trim=0.cm 0.cm 1.cm 0.cm, clip=true]{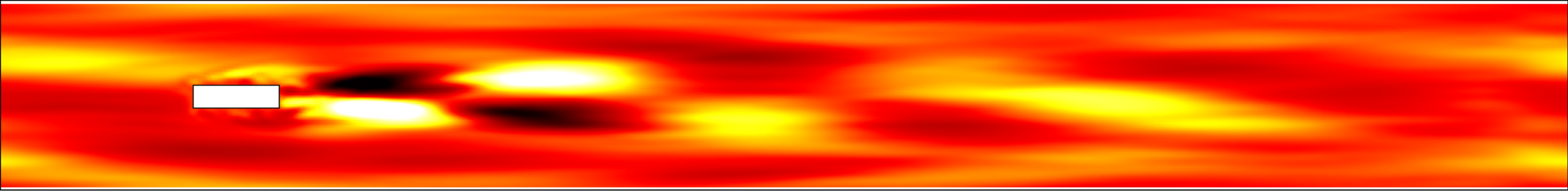}
    \end{adjustbox}
    \begin{adjustbox}{width=14cm}
        	      k=11\hspace{1.5mm}\includegraphics[width=6.3cm,trim=0.cm 0.cm 1.cm 0.cm, clip=true]{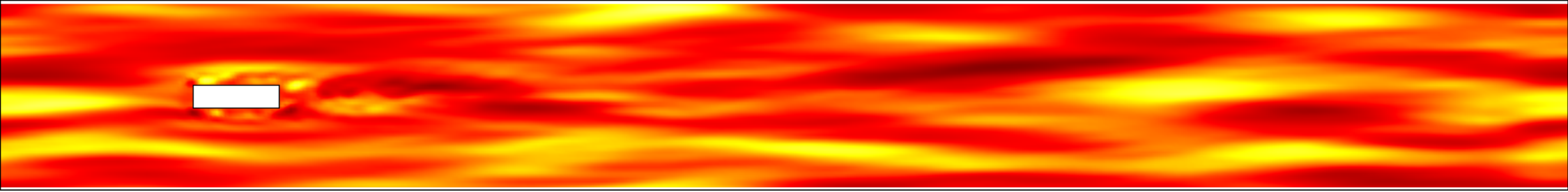}
        \hspace{10.mm}k=8\hspace{2.5mm}	\includegraphics[width=6.3cm,trim=0.cm 0.cm 1.cm 0.cm, clip=true]{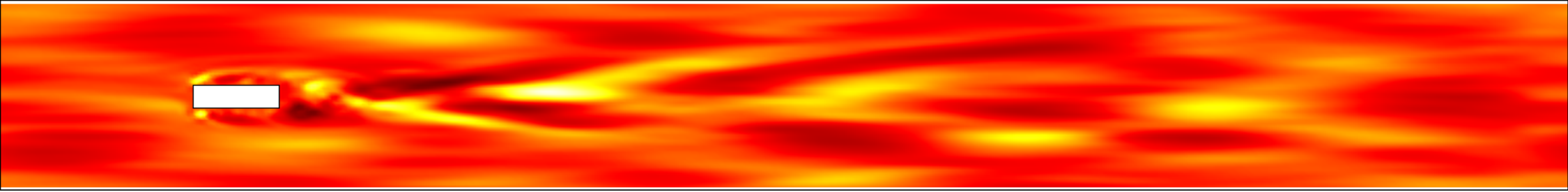}
    \end{adjustbox}
    \begin{adjustbox}{width=14cm}
        	      k=16\hspace{1.5mm}\includegraphics[width=6.3cm,trim=0.cm 0.cm 1.cm 0.cm, clip=true]{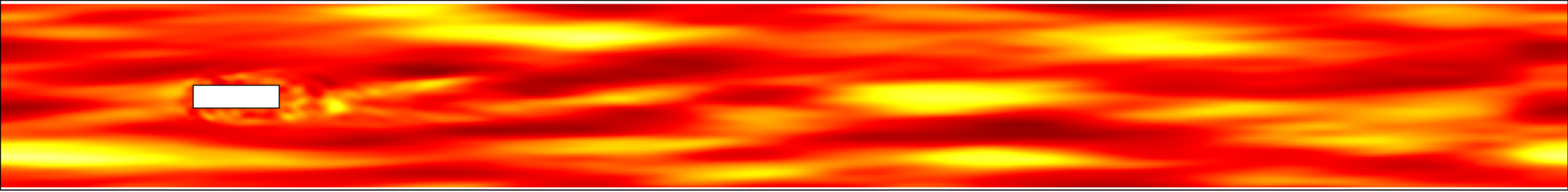}
        \hspace{10.mm}k=13\hspace{1.5mm}\includegraphics[width=6.3cm,trim=0.cm 0.cm 1.cm 0.cm, clip=true]{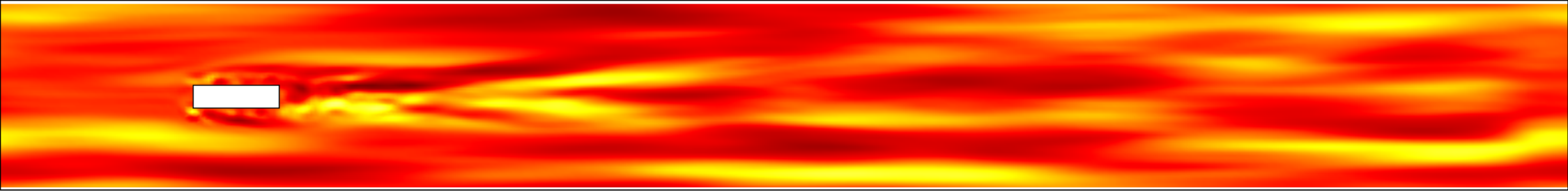}
    \end{adjustbox}
    \begin{adjustbox}{width=14cm}
        	      k=18\hspace{1.5mm}\includegraphics[width=6.3cm,trim=0.cm 0.cm 1.cm 0.cm, clip=true]{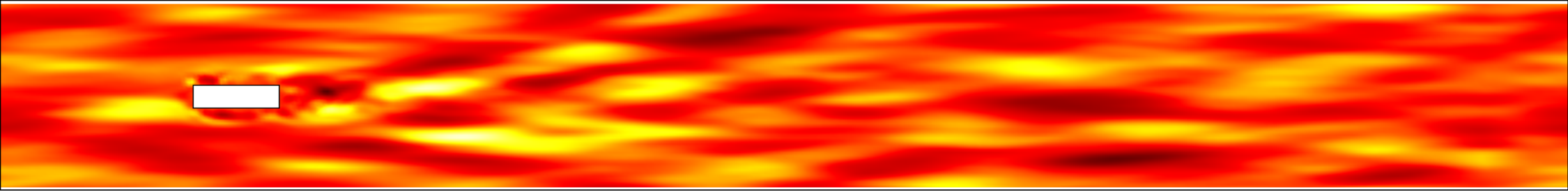}
        \hspace{10.mm}k=16\hspace{1.5mm}\includegraphics[width=6.3cm,trim=0.cm 0.cm 1.cm 0.cm, clip=true]{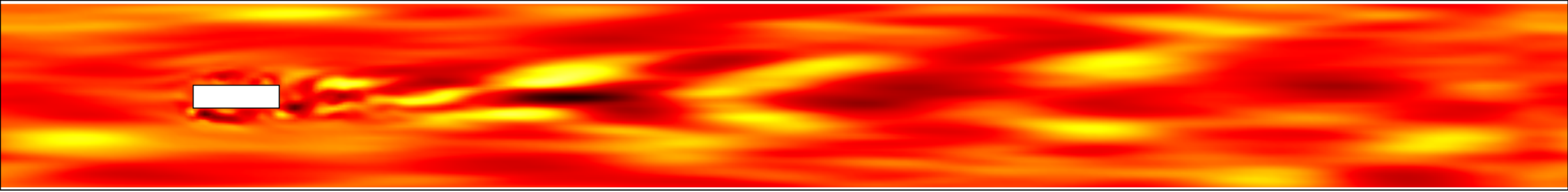}
    \end{adjustbox} \\
      c) \hspace{40mm}  d)    
 \end{center}
  \caption{Streamwise velocity POD mode contours on a horizontal plane sectioning the box through the center for modes at constant relative energy levels: a) $\Delta\theta = 0.2$ K, $U_g=8$ m/s; b) $\Delta\theta = 0.2$ K, $U_g=18$ m/s; c) $\Delta\theta = 0.4$ K, $U_g=8$ m/s; d) $\Delta\theta = 0.4$ K, $U_g=18$ m/s. The ordering of the contours corresponds to the rows in table \ref{mode_compare}.}
  \label{POD_u}
\end{figure}

Next, contour plots of several streamwise velocity POD modes are compared at constant relative energy levels. The horizontal lines in figure \ref{mode_energy}(a) highlight the mode numbers across all four cases, within the first twenty modes, that have constant relative energy levels, $\frac{\lambda^k}{\lambda^1}$, which are shown in Table \ref{mode_compare}. The contours of the streamwise velocity POD modes shown in figure \ref{POD_u} correspond to the relative energy levels in table \ref{mode_compare}. For example, the second contour from the top in figures \ref{POD_u}(a), \ref{POD_u}(b), \ref{POD_u}(c), and \ref{POD_u}(d) corresponds to the second relative energy level in table \ref{mode_compare}. For the thermal stratification $\Delta\theta = 0.2$ K and wind velocity $U_g=8$ m/s contours shown in figure \ref{POD_u}(a), the modes of the first four energy levels are of the weak combined helical-symmetric type that shows no significant effect from the box. The fifth energy level mode shows the behavior of the dominant helical type, although in a weaker form. The contours of the modes at a thermal stratification $\Delta\theta = 0.2$ K and wind velocity $U_g=18$ m/s in figure \ref{POD_u}(b) show a clear dominant helical mode at the third energy level, while the modes of the first two energy levels are largely unaffected and the modes of the last two energy levels show a  stronger combined helical-symmetric type behavior. Figure \ref{POD_u}(c) shows the mode contours for the thermal stratification $\Delta\theta = 0.4$ K and wind velocity $U_g=8$ m/s. Clear dominant helical modes are seen at the highest two energy levels. The next two energy levels display modes with a weak combined helical-symmetric behavior and the mode at last energy level exhibits weaker helical behavior at a larger streamwise wavenumber than the modes at the first two energy levels. The contours of the modes at a thermal stratification $\Delta\theta = 0.4$ K and wind velocity $U_g=18$ m/s are seen in figure \ref{POD_u}(d). The dominant helical modes are still seen at the first two highest energy levels, except stronger, while the last three energy levels all display clearer dominant helical type mode behavior. 

\begin{table}[htpb]
 \begin{center}
  \begin{adjustbox}{width=7.8cm}
   \begin{tabular}{| c || c | c | c | c |} \hline
       \multirow{3}{*}{$\lambda^k$/$\lambda^1$} & 	\multicolumn{4}{ c| }{Mode Number}\\ \cline{2-5}
		&	$\Delta \theta = 0.2$ K	&	$\Delta \theta = 0.2$ K	&	$\Delta \theta = 0.4$ K	& $\Delta \theta = 0.4$ K \\ 
       		&	$U_g = 8$ m/s	&	$U_g = 18$ m/s	&	$U_g = 8$ m/s	& $U_g = 18$ m/s \\ \hline\hline
       $1.0$	&	$1$	&	$1$	&	$1$	&	$1$ \\ \hline
       $0.981$	&	$2$	&	$2$	&	$2$	&	$2$ \\ \hline
       $0.584$	&	$5$	&	$5$	&	$11$	&	$8$ \\ \hline
       $0.496$	&	$9$	&	$7$	&	$16$	&	$13$ \\ \hline
       $0.466$	&	$11$	&	$8$	&	$18$	&	$16$ \\ 
\hline
   \end{tabular}
  \end{adjustbox}
    \caption{POD mode numbers ($k$) at constant relative energy levels ($\frac{\lambda^k}{\lambda^1}$).}
  \label{mode_compare}
 \end{center}
\end{table}

Increasing the wind velocity causes the dominant helical modes to appear at higher energy levels. This is seen in the $\Delta\theta = 0.2$ K cases where there is a weaker helical mode at the fifth energy level when $U_g=8$ m/s and a stronger helical mode at the third energy level when $U_g=18$ m/s and also in $\Delta\theta = 0.4$ K cases where helical modes appear at the last three energy levels when $U_g=18$ m/s, but only at the last energy level when $U_g=18$ m/s. The $\Delta\theta = 0.4$ K cases are also interesting because clear helical modes are present at the highest two energy levels for both wind velocities, with the modes associated with the higher velocity appearing stronger than the modes associated with the lower velocity at a constant energy level. The appearance of helical modes at higher energy levels with increasing wind velocity and the strength of helical modes at constant energy levels increasing with increasing wind velocity can be related back to the increase of turbulence intensity and vertical turbulent momentum flux in the wake found in figures \ref{varU_02} and \ref{varUW_02}. Since the helical modes are the dominant modes, as the wake becomes more turbulent these helical modes contain more of the turbulent energy. Increasing the thermal stratification causes a similar effect as increasing the the wind velocity; the dominant helical modes appear at higher energy levels at higher thermal stratifications. Looking back at figures \ref{varU_10} and \ref{varUW_10}, the profiles in the wake of the turbulence intensity and the vertical turbulent momentum flux were only slightly affected by varying the thermal stratification. This can be interpreted as the increase in thermal stratification concentrating the turbulent energy in the higher energy levels, which manifests as more helical modes at higher energy levels, while leaving the overall statistics largely unaffected. 

\begin{figure}
 \begin{center}
    \begin{adjustbox}{width=12cm}
        	      k=6\hspace{2.5mm} \includegraphics[width=12.5cm,trim=0.cm 0.cm 1.cm 0.cm, clip=true]{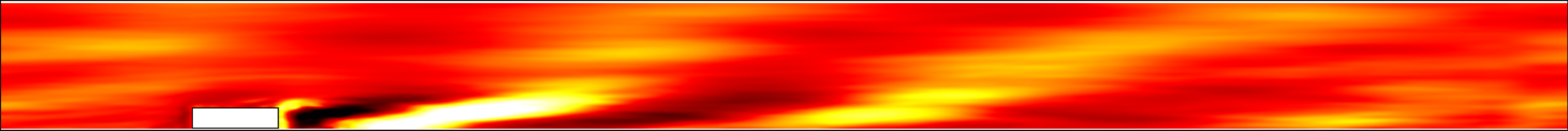}
    \end{adjustbox} \\
    a) \\ \vspace{2.5mm}
    \begin{adjustbox}{width=12cm}
        	      k=5\hspace{2.5mm} \includegraphics[width=12.5cm,trim=0.cm 0.cm 1.cm 0.cm, clip=true]{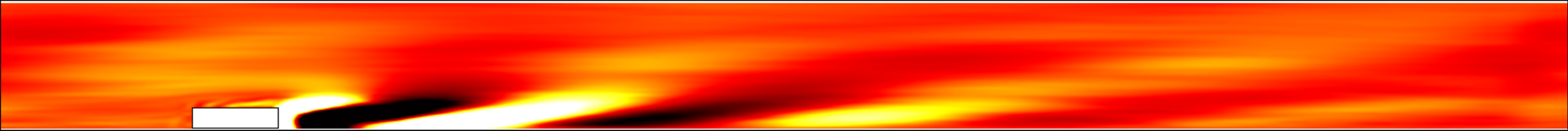}

    \end{adjustbox} \\
    b) \\ \vspace{2.5mm}   
   
    \begin{adjustbox}{width=12cm}
        	      k=1\hspace{2.5mm} \includegraphics[width=12.5cm,trim=0.cm 0.cm 1.cm 0.cm, clip=true]{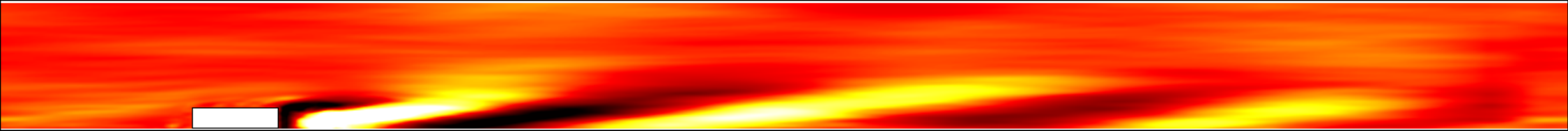}
        	      
    \end{adjustbox} \\
    c) \\ \vspace{2.5mm}
    \begin{adjustbox}{width=12cm}
        	      
        	      k=1\hspace{2.5mm} \includegraphics[width=12.5cm,trim=0.cm 0.cm 1.cm 0.cm, clip=true]{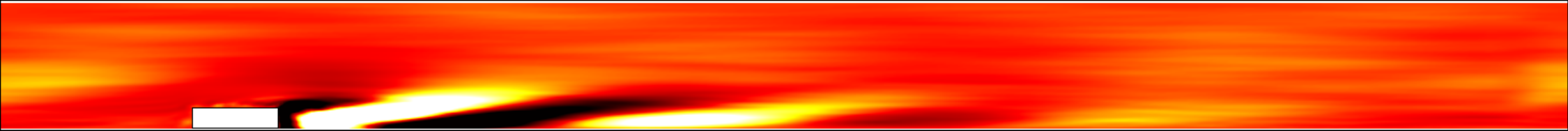}
    \end{adjustbox} \\
    d)  
 \end{center}
  \caption{Streamwise velocity POD mode contours on a vertical plane sectioning the box through the center for modes of a constant streamwise wavenumber: a) $\Delta\theta = 0.2$ K, $U_g=8$ m/s; b) $\Delta\theta = 0.2$ K, $U_g=18$ m/s; c) $\Delta\theta = 0.4$ K, $U_g=8$ m/s; d) $\Delta\theta = 0.4$ K, $U_g=18$ m/s.}
  \label{POD_v}
\end{figure}

While the previous comparisons of the POD modes were made at constant relative energy levels, next we focus on comparisons of modes at constant streamwise wavenumber. Contours of the spanwise velocity POD modes are presented in figure \ref{POD_v}. These contours show the first occurrence of a dominant helical mode at a constant streamwise wavenumber for each of the four cases. For a constant streamwise wavenumber and comparing at a constant thermal stratification, figures \ref{POD_v}(a) and \ref{POD_v}(b) for $\Delta\theta = 0.2$ K and figures \ref{POD_v}(c) and \ref{POD_v}(d) for $\Delta\theta = 0.4$ K, it is seen that the strength of the helical mode increases with increasing wind velocity. This agrees with the trends seen above.  Now, comparing at a constant wind velocity, figures \ref{POD_v}(a) and \ref{POD_v}(c) for $U_g=8$ m/s and figures \ref{POD_v}(b) and \ref{POD_v}(d) for $U_g=18$ m/s, it is seen that the mode inclination angle (with respect to the bottom boundary) decreases a small amount as thermal stratification increases. The increased thermal stratification forces the modes to spread more in the horizontal directions and less in the vertical direction. This is consistent with the small decrease in vertical turbulent momentum flux around the box height seen in figure \ref{varUW_10}; increasing the thermal stratification ``contains'' the wake closer to the wall.

The Strouhal number based on the length of the box for the dominate helical modes for each case is shown in table \ref{Strouhal}. It can be observed in figure \ref{POD_u} that an increase in wind velocity causes the wavelength of the of the helical mode to decrease, which in turn causes an increase in Strouhal number.  The decrease in wavelength occurs because the helical modes contain more higher energy structures at the higher wind velocity. Also from figure \ref{POD_u}, the wavelength of the helical mode increases as the thermal stratification increases, which yields a decrease in Strouhal number. This relates back to the increased thermal stratification limiting the vertical spread of the modes. The smaller thermal stratification allows for a larger increase in the Strouhal number as the wind velocity increases because the modes are allowed to spread more in the vertical direction. 

\begin{table}[htpb]
 \begin{center}
  \begin{tabular}{| c || c | c |} \hline
       $St_{l_{box}}$          & $U_g = 8$ m/s 	    & $U_g = 18$ m/s \\	\hline \hline
       $\Delta\theta = 0.2$ K  & $0.26$             & $0.31$	      \\	\hline
       $\Delta\theta = 0.4$ K  & $0.23$             & $0.25$  	      \\
\hline
  \end{tabular}
    \caption{Box length Strouhal number for the dominate helical modes.}
  \label{Strouhal}
 \end{center}
\end{table}

\section{Conclusions}

Large eddy simulations were used to study the dynamics of the wake created by a large bluff body in a high Reynolds number stably thermally-stratified boundary layer. The boundary conditions at the surface of the body were modeled using a direct forcing immersed boundary method. Since the LES tool uses a pseudo-spectral discretization, requiring the use of periodic boundary conditions in the horizontal directions, the inflow conditions were handled via a concurrent precursor simulation method aimed at generating realistic turbulent inflow conditions. This allowed the simulation of a single row of boxes (the periodic condition in the spanwise direction was applied). The three-dimensional snapshot POD method was used to study the structure of the wake generated by the body. Two different thermal stratifications and two wind velocities were considered in the analysis. The computational method was validated using experimental results from a wind tunnel flow around a wall-mounted cube.

A large bluff body in a stable boundary layer was then considered, and results consisting of vertical profiles of vertical turbulent momentum and heat fluxes and turbulence intensity were reported and discussed. The turbulent statistics were seen to be significantly affected by the presence of the bluff body. The thermal stratification variation was found to have just a slight effect on all statistical quantities, but the larger wind velocity significantly affected all the vertical profiles when compared to the smaller wind velocity. From the POD analysis, two distinct POD modes were observed in the wake. These two classifications of modes were labeled helical and symmetric and had different streamwise wavenumbers. The helical mode was found to be dominant; the more turbulent the wake became, the more helical modes were found at the highest energy levels. Contrary to the slight effect seen in the statistical quantities, the distribution of helical modes at the highest energy levels was significantly affected by varying the thermal stratification.

Future work will focus on an improvement of the current concurrent precursor method in order to provide a more natural transition from the main to the precursor flow at the outlet of the main domain. In addition, the concurrent precursor method will also be compared with a synthetic eddy method, also in development, in an attempt to improve the computational speed while still generating realistic turbulent inflow conditions because the current concurrent precursor method is computationally expensive.

\section*{Acknowledgement}

The authors would like to thank Dr. Claire VerHulst for Johns Hopkins University for her help in regard to the POD analysis.



\end{document}